\documentclass[3p]{elsarticle}
\graphicspath{ {./media/} }
\usepackage{hyperref}
\usepackage{float}
\usepackage{verbatim} %comments
\usepackage{apalike}
\restylefloat{figure}
\restylefloat{table}
\usepackage{amsmath}
\usepackage{hyperref}
\usepackage{algorithmic}
\usepackage{array}
\usepackage{fixltx2e}
\usepackage{stfloats}
\usepackage{amsfonts}
\usepackage{multicol}
\usepackage{float}
\usepackage{booktabs}
\usepackage{soul}

\DeclareMathOperator{\sign}{sign}
\newcommand{\ra}[1]{\renewcommand{\arraystretch}{#1}}

\journal{/ to be submitted}

\biboptions{authoryear}

\begin{document}
\begin{frontmatter}

% UNCOMMENT WHEN SUBMITTED TO ELSEVIER 

%\begin{titlepage}
%\begin{center}
%\vspace*{1cm}
%
%\textbf{ \large Antifragile Control Systems: The case of an oscillator-based network model of urban road traffic dynamics}
%
%\vspace{1.5cm}
%
%% Author names and affiliations
%Cristian~Axenie$^{a,b}$(cristian.axenie@huawei.com)
%
%\hspace{10pt}
%
%\begin{flushleft}
%\small  
%$^a$ Huawei Munich Research Center, Riesstraße 25, 80992 Munich, Germany \\
%$^b$ Audi Konfuzius--Institut Ingolstadt Laboratory, Technische Hochschule Ingolstadt, Esplanade 10, 85049 Ingolstadt, Germany
%
%\vspace{1cm}
%\textbf{Corresponding Author:} \\
%Cristian~Axenie \\
%Intelligent Cloud Technologies Laboratory, Huawei Munich Research Center, Riesstraße 25, 80992 Munich, Germany \\
%Phone: +49(0) 174 321 1689 \\
%Email: cristian.axenie@huawei.com
%
%\end{flushleft}        
%\end{center}
%\end{titlepage}

\title{Antifragile Control Systems: The case of an oscillator-based network model of urban road traffic dynamics}

\author[label1,label2]{Cristian~Axenie\corref{cor1}}
\ead{cristian.axenie@huawei.com; cristian.axenie@gmail.com}

\author[label1]{Margherita~Grossi}

\cortext[cor1]{Corresponding author.}
\address[label1]{Intelligent Cloud Technologies Laboratory, Huawei Munich Research Center, Riesstraße 25, 80992 Munich, Germany}
\address[label2]{Audi Konfuzius-Institut Ingolstadt Laboratory, Technische Hochschule Ingolstadt, Esplanade 10, 85049 Ingolstadt, Germany}

\begin{abstract}

Urban road traffic is a highly nonlinear process which continuously evolves under uncertainty. Short-term sporadic events can determine large changes in traffic flow which propagate in both space and time across an urban road network. Existing traffic control systems only possess a local perspective over the multiple scales of traffic evolution, namely the intersection-level, the corridor-level, and the region-level respectively. Aggregating such local views of traffic evolution over a large region is what current control systems try by using traffic models. Yet, capturing uncertainty under such complex spatio-temporal interactions is a very difficult problem and we often experience how fragile such systems are in reality. Fortunately, despite its complex mechanics, traffic is described by various periodic phenomena. Workday flow distributions in the morning and evening commute times can be exploited to make traffic adaptive and robust to disruptions. Additionally, controlling traffic is also based on a periodic process, choosing the phase of green time to allocate to opposite directions right of pass and complementary red time phase for adjacent directions. In our work, we harmonize this perspective and consider a novel system for road traffic control based on a network of interacting oscillators. Such a model has the advantage of capturing both temporal and spatial interactions of traffic light phasing, as well as the network-level evolution of the traffic macroscopic features (i.e. flow, density). We demonstrate that periodicity exploiting closed-loop control systems, which are capturable by such a model, have the potential to tame periodic dynamics and are robust to the inherent uncertainty in traffic evolution. In this study, we propose a new realization of the antifragile control framework to control a network of interacting oscillator-based traffic light models to achieve region-level flow optimization. We demonstrate that antifragile control can capture the volatility of the urban road environment and the uncertainty about the distribution of the disruptions that can occur. We complement our control-theoretic design and analysis with experiments on a real-world setup to comparatively discuss the benefits of an antifragile design for traffic control.
\end{abstract}


\end{frontmatter}

\section{Introduction}

\begin{figure} 
\centering
\includegraphics[scale=0.25]{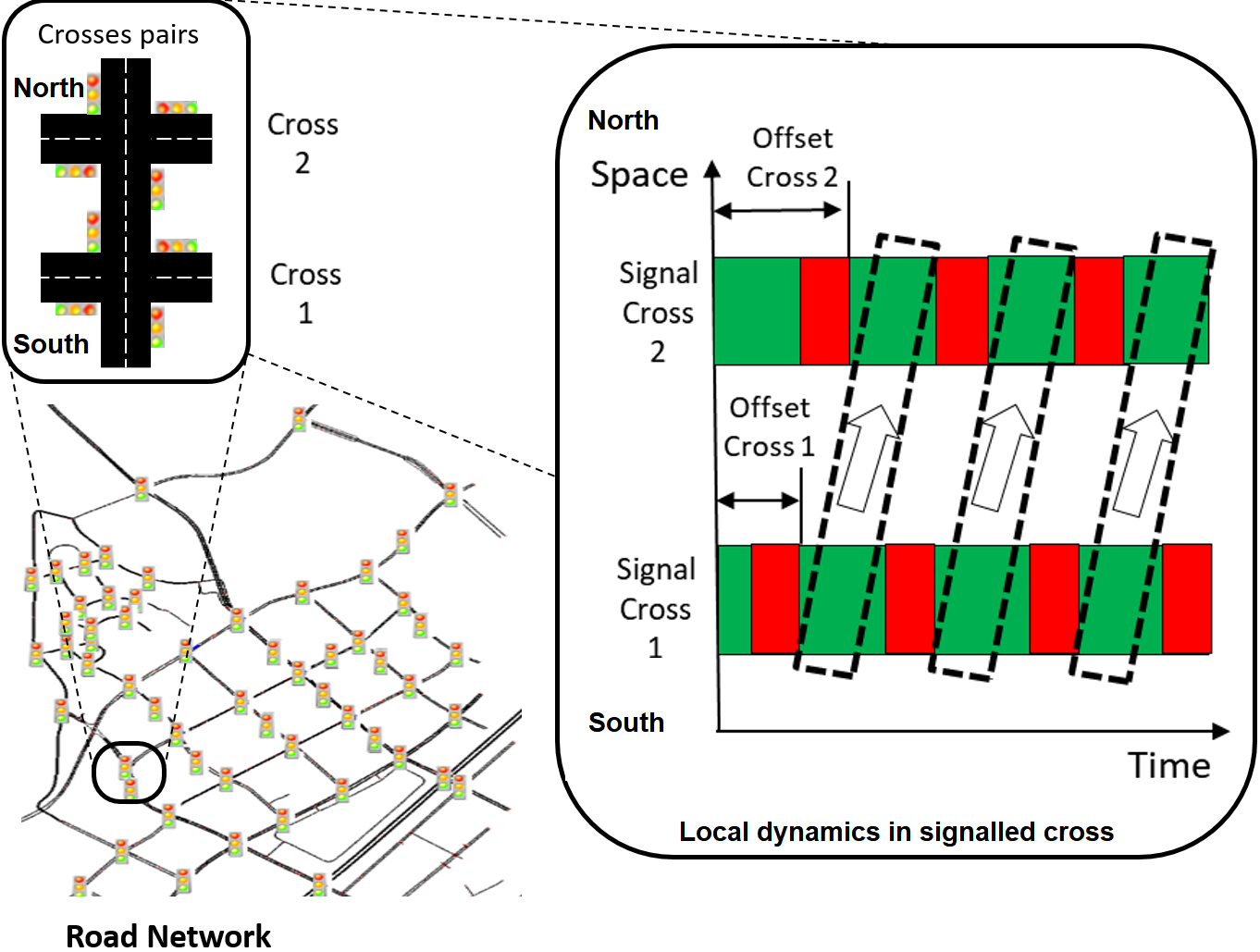}
\caption{Traffic control signaling. The fundamental dynamics in a pair of signalized crosses is described by the space-time diagram of the platoon of cars passing from South $\Leftrightarrow$ North for the duration of the green light signal. The periodic signal behavior ensures that adjacent directions (i.e. West $\Leftrightarrow$ East transit) are allowed to pass through (not shown in space-time). The control algorithm should compute the green and red time signals while taking into account an offset to enable safe sequencing of driving directions.}
\label{fig1}
\end{figure}

Urban road traffic congestion is still resistant to straightforward solutions, given the strong impact it has on infrastructure as well as the economic and social dimensions of life in large cities. Traffic modelling, control, and optimization remain among the hardest problems across disciplines where only "incremental" research pushes a slow progress, as emphasized in the review of \cite{van2015genealogy}. Considering actual technology instantiations, such as SCOOT from \cite{hunt1982scoot}, SCATS from \cite{lowrie1990scats}, PRODYN from \cite{henry1984prodyn}, OPAC from \cite{gartner1983opac}, RHODES from \cite{mirchandani1998rhodes}, or LISA from \cite{LISA}, adaptive traffic signal control systems use a simple model of traffic dynamics and feed it with sensor-based vehicles detection in order to optimize signal timings. 
Subsequently, detections from multiple crosses, are aggregated into a central system, which models the flow of traffic in the area. Finally, the underlying traffic model is used to adapt the phasing of the traffic light signals in accordance with the flow of traffic, thus minimizing unnecessary green phases and allowing the traffic to flow most efficiently. For a basic depiction of the space-time dynamics of signalized road traffic, we refer the reader to Figure~\ref{fig1}. The focus of our work is to provide a control framework that can handle uncertainty which prevails in urban traffic dynamics and inherent anomalies over regular traffic patterns.

\subsection{Traffic control}

In principle, closed-loop road traffic control could scale to city-level, in an ideal scenario, when all the temporal and spatial interactions among the crosses are known and precisely modelled. But, this is never the case and additional dimensions emerge. For instance uncertainty in the weekly patterns of traffic load on fixed capacity arteries (see work of \cite{chen2010modeling}), volatility of hourly capacity in multi-lane streets in the city center or during sport activities (see work of \cite{ossenbruggen2012congestion}), and variability of the external traffic entering the city or heavy rain during late autumn (see work of \cite{ko2006variability}) are clear components of traffic dynamics. Such components make traffic contexts fragile towards degenerating into congestion. The problem lies in the fact that such phenomena are hard to model, predict, and control. 

Technically, the core differentiating aspect among the existing systems is their underlying traffic model, in other words, the dynamics of traffic they capture and how this model handles the inherent uncertainty, volatility, and variability of the captured variables. For instance, based on large amounts of high-resolution field traffic data, the work of \cite{hu2013arterial} used the conditional distribution of the green signal times and traffic demand to improve urban flow. However, such data were expensive to acquire and the statistical method couldn't handle long-tail events (e.g. daily traffic patterns fluctuations on weekends vs. week days). Using a relatively simple model to predict arrivals at coordinated signalled crosses, the work of \cite{day2020optimization} assumes nearest-neighbor interactions between signals and uses a linear superposition of distributions to optimize traffic lights phase duration. Despite finding the optimal coordination, the algorithm couldn't handle unpredictable changes to platoon shapes (i.e. occasionally caused by platoon splitting and merging) or unpredictably saturated conditions (i.e. traffic jams, accidents). Such limited adaptation capabilities in the face of uncertainty and disruptions, which can propagate in time and space throughout the system and lead to heavy congestion, make the system fragile. 

The main goal of this study is to introduce the application of antifragile control to traffic control and optimization under uncertainty, volatility, and variability. As coined in the book of Taleb \cite{taleb2012antifragile}, antifragility is a property of a system to gain from uncertainty, randomness, and volatility, opposite to what fragility would incur. An antifragile system's response to external perturbations is beyond robust, such that small stressors can strengthen the future response of the system by adding a strong anticipation component. For the closed-loop system we employ the model introduced by \cite{axenie2021obelisc} that uses oscillator-based dynamics modelling. Such an approach has the advantage of capturing both the periodicity in daily/seasonal traffic patterns and the periodicity of the local signal timings, and as shown in \cite{axenie2021obelisc} is robust and efficient at scale. We now propose an alternative control mechanism, based on the antifragile control framework introduced by \cite{axenie2022antifragile} and built on top of the principles in the seminal work of \cite{taleb2013mathematical}.

\subsection{Oscillator-based network modelling and interaction dynamics}

Oscillator-based modelling and control is an approach emerging from physics that proves to be a plausible application in traffic control. In a very good review and perspective, the study of \cite{chedjou2018review} introduced the formalism of oscillator-based traffic modelling and control. Despite the strong mathematical grounding, the proposed approach was static, in that it removed all convergence and dynamics of the oscillator-based model by replacing it with the steady-state solution. Such an approach proves its benefits at the single cross level, as the authors claim, but will fail in large-scale heterogeneous road networks (i.e. non-uniform road geometry, disrupted traffic patterns, volatility of traffic load, uncertainty of weather conditions). The approach we take in the current study is to design a closed-loop antifragile control system in which a novel variable structure sliding mode controller (see \cite{utkin2008sliding}) is designed for a nonlinearly-coupled oscillators model based on the model of \cite{strogatz2000kuramoto}. We demonstrate through our experiments that the oscillator-based model with antifragile control can stabilize in a plausible solution of signal timing under dynamical demand changes based on measurement of local traffic data. A similar oscillator-based model approach was used in the work of \cite{nishikawa2004dynamics} and later in the work of \cite{fang2013matsuoka} as area-wide signal control of an urban traffic network. Yet, due to their complex-valued dynamics and optimization, the systems could not capture both the spatial and temporal correlations under a realistic computational cost for real-world deployment.

\subsection{Fragility-robustness-antifragility continuum in traffic control}

\begin{figure} 
\centering
\includegraphics[scale=0.3]{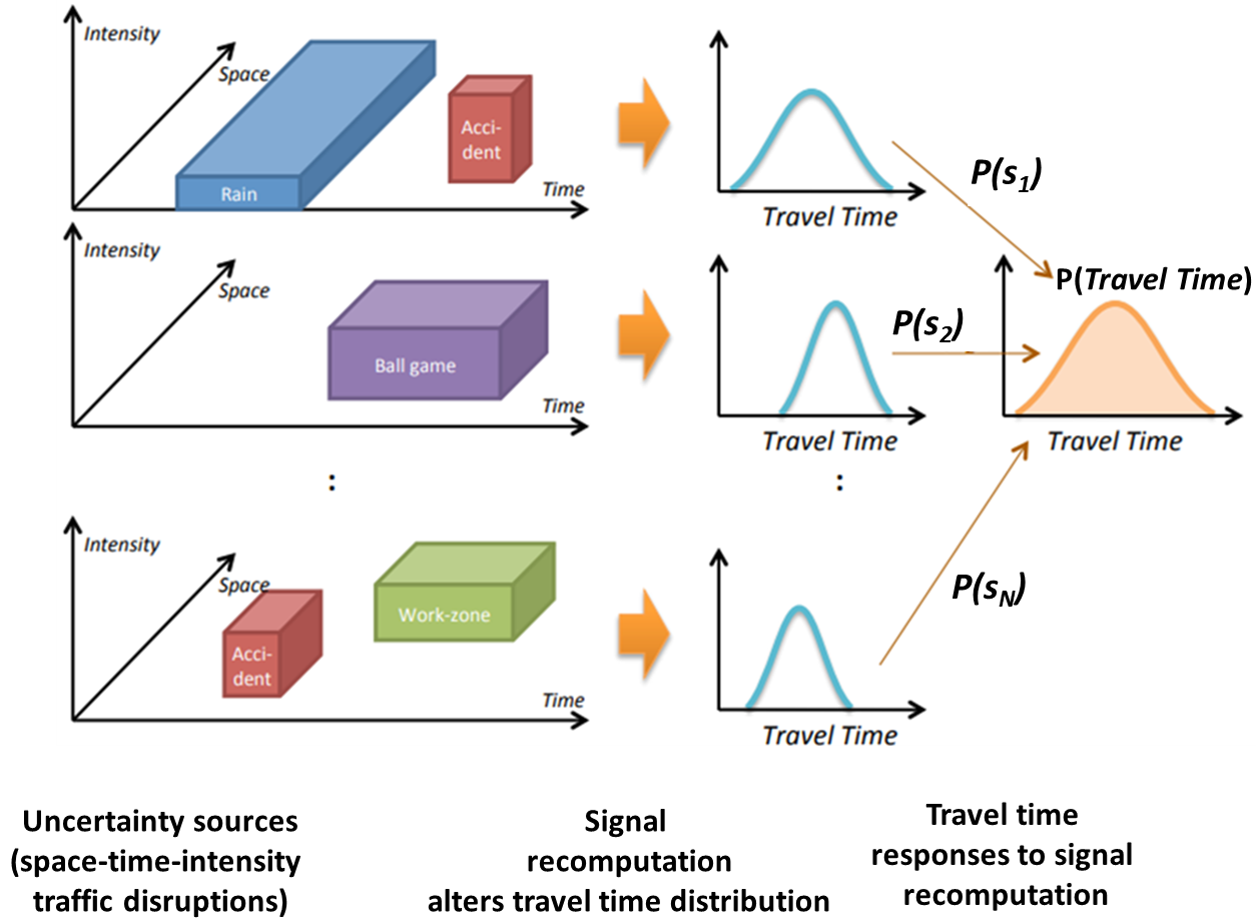}
\caption{Uncertainty, volatility and variability in traffic control, when considering travel time over an itinerary. Space-time-intensity of various sources of uncertainty in traffic (e.g. weather, accidents, social events, infrastructure availability) and the traffic signal re-computation that alters the distribution of travel time. The overall (whole itinerary) travel time distribution change as response to signal re-computation.}
\label{fig2}
\end{figure}

Traffic dynamics is highly nonlinear and sensitive to multiple sources of uncertainty. Here we go beyond uncertainty in capturing the real dynamics of traffic through a model (i.e. structured/parametric uncertainties) and consider the changes that disruptions, such as weather, accidents, social events, and infrastructure availability (i.e. unstructured uncertainties, or un-modelled dynamics), induce in the overall flow of cars. The uncertainty, volatility, and variability inherent in such disruptions are described by a stochastic evolution in the space-time-intensity reference system. The compound effect of such disruptions (typically additive in nature) reflects itself in computed measures of quality of traffic, for instance travel time for cars over a certain itinerary. Such unstructured uncertainties determine traffic signal re-computations that, subsequently, alter the shape of the travel time distribution and, hence, the overall travel time distribution -- as depicted in Figure~\ref{fig2}.

\begin{figure} [!ht]
\centering
\includegraphics[scale=0.3]{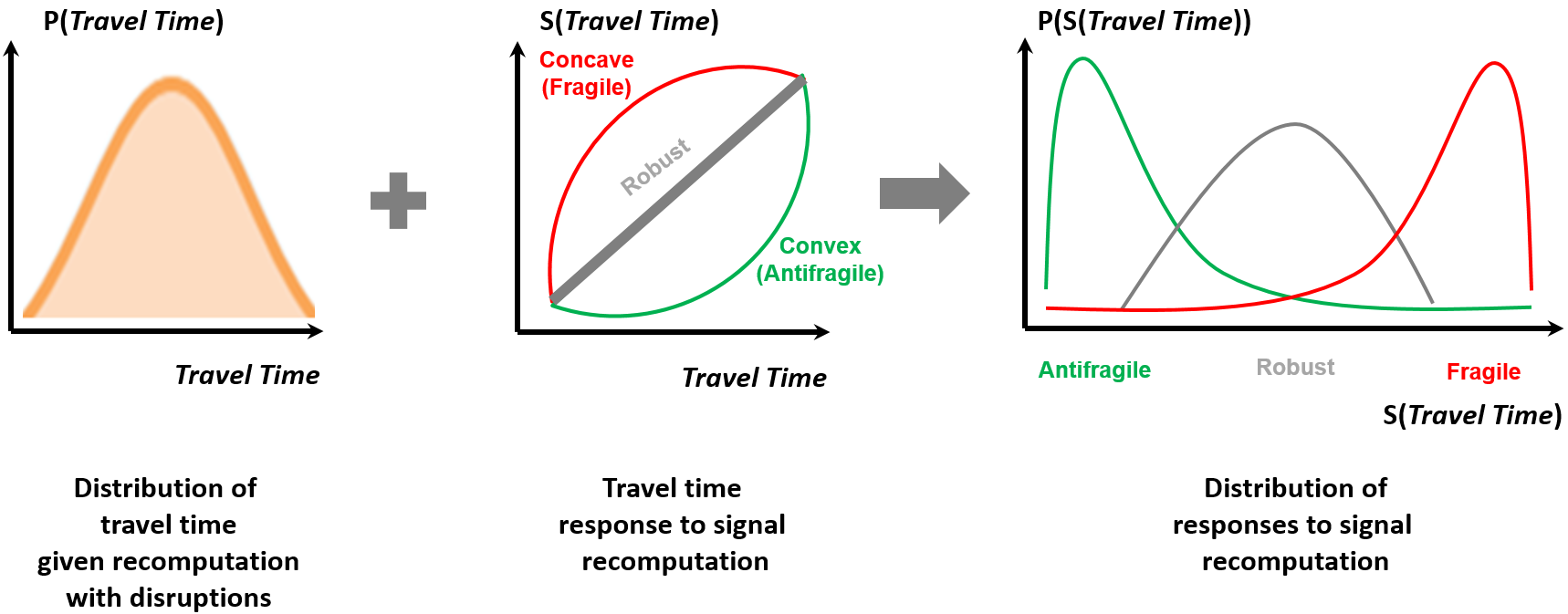}
\caption{Mapping from uncertainty effects on travel time $P(Travel Time)$ to the fragile-robust-antifragile spectrum under signal re-computation. Signal re-computation response of travel time changes shape $S(Travel Time)$ can vary based on the closed-loop control variables and push the system towards a different region of the response to signal re-computation distribution $P(S(Travel Time))$. Due to the periodic nature of the traffic light signalling the shape of $S(Travel Time)$ can be convex (i.e. antifragile behavior), concave (i.e. fragile behavior), linear (i.e. robust behavior), or even mixed convex--concave (i.e. non-stationary behavior).}
\label{fig3}
\end{figure}

Going further with our travel time example, the change in the shape of the distribution of the travel time can be described through the actual type of response (i.e. the shape) to the space-time-intensity characteristics of the uncertainty. We can then describe the distribution of such responses to uncertainty-triggered signal re-computation with respect to the travel time itself, as shown in Figure~\ref{fig3}. As also show in the excellent work of \cite{taleb2022working}, if we combine the reaction to uncertainties, or the uncertainty response (i.e. signal re-computation) $S(Travel Time)$, with the distribution of the travel time $P(Travel Time)$, we can describe the probability distribution of the signal re-computation $P(S(Travel Time))$. The core idea is that we can vary the parameters of the signal re-computation such that the shape of $S(Travel Time)$ changes to handle the changes in $P(Travel Time)$ given uncertainty.

\begin{figure} [!ht]
\centering
\includegraphics[scale=0.3]{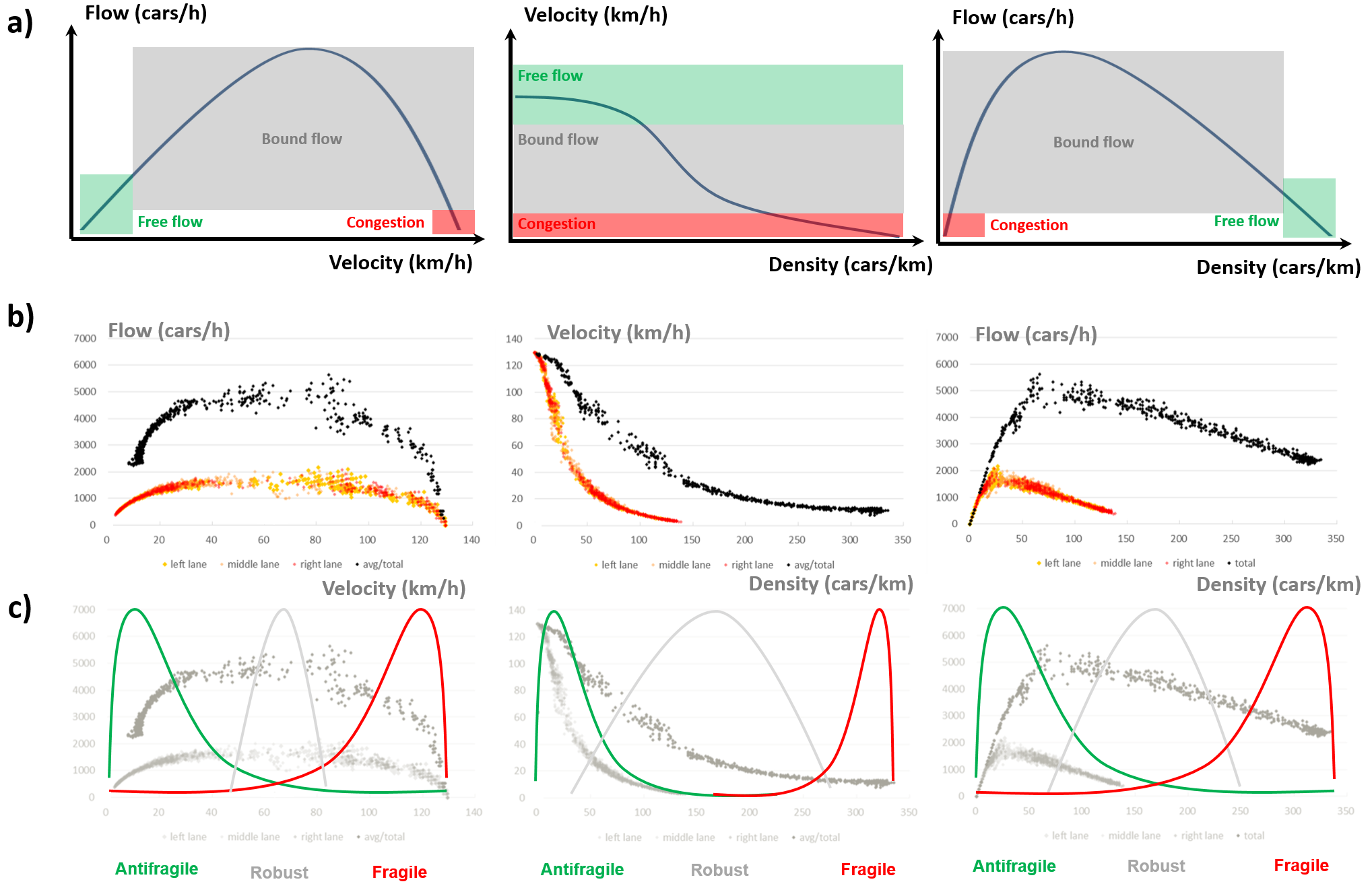}
\caption{Mapping macroscopic fundamental diagrams (MFD) to the fragile--robust--antifragile spectrum in road traffic control under uncertainty. a) Analytic form of the MFD and the traffic regimes characteristics. b) Real MFD extracted from a real-world dataset on a highway segment with three lanes. c) The fragile--robust--antifragile distributions shapes based on the mapping of MFD on the real data curves. The shapes of the distribution in the loss and the gain regions (on the MFD) are then matched using Taleb's heuristic (see \cite{taleb2013mathematical}) to the fragile--robust--antifragile continuum.}
\label{fig4}
\end{figure}

The unique configuration of space-time-intensity of the uncertainty actually determines the critical points where the traffic dynamics would not be able to compensate for uncertainty, and become fragile. In traffic engineering, the spectrum of macroscopic behaviors is captured through macroscopic fundamental diagrams (MFD). As systematically introduced in the work of \cite{li2011fundamental}, MFD describe bivariate equilibrium relationships of traffic flow, density, and speed that can provide a versatile mapping to the fragile-robust-antifragile spectrum. The alignment among the two spectra is depicted in Figure~\ref{fig4}. Here, we consider the theoretical MFD, the definition of the three possible equilibrium states--free flow, bound flow, and congestion--and their placement in the velocity--density, flow--velocity, and flow--density characteristics (see Figure~\ref{fig4} a).

As one can see in Figure~\ref{fig4} b, the real MFD indicate that both capacity drop and concave-–convex MFD shapes abound in practice. In this example, the detector data used to compute the MFD in Figure~\ref{fig4} b is considering a highway scenario from the open-source SUMMER-MUSTARD dataset of \cite{cristian_axenie_2021_5025264} available on Zenodo\footnote{Data available at: \url{https://zenodo.org/record/5025264}}. Important to note that the analysis is based on a highway scenario, so there's no traffic signal control. We chose to first describe the behavior mapping when traffic dynamics evolve without control. More precisely, we plot the MFD for a three lane road segment (excluding the high-occupancy lane) in China over a day. We can map the MFD regions, and implicitly traffic dynamics, to the fragile--robust--antifragile spectrum, as shown in Figure~\ref{fig4} c. We use the taxonomy provided \cite{taleb2012antifragile} and the mathematical identification heuristic in \cite{taleb2013mathematical} to map the loss and gain domains in terms of the distribution of points in the road traffic MFD in Figure~\ref{fig4} a, b. 

Considering the flow--velocity characteristic, the captured shape from the data follows the analytic shape and we can, hence, identify a direct mapping. Here the free flow region corresponds to a fat tail in the gain domain (i.e. increasing velocity with sub-linearly increasing flow) and a thin tail in loss domain (i.e. high velocity and high flow), practically corresponding to congestion region of the MFD. The robust behavior materializes in the region in which moderate velocity yields the maximal flow of cars (i.e. bound flow), and with a distribution characterized by a thin tail in loss domain and a thin tail in gain domain.

When considering the velocity--density characteristic we can see that the real data matches the analytic shape with a large robust region covering the density span from 75 to 275 cars/km where the velocity decreases towards a creeping regime where the congestion forms. The antifragile region distribution of the characteristic is capturing the high-velocity average density of the MFD where uncertainty can only push the system right across the robust region (i.e. thin tail in loss region and this tail in the gain region). At the other end of the spectrum, we have the fragile region where congestion forms with a fat tail forming while the velocity decreases at very high density, where congestion is already in place.

The flow--density characteristic already starts demonstrating the limitations, or the fragile region, of traffic dynamics without control. More precisely, in Figure~\ref{fig4} b, we see that the flow of cars doesn't decrease at high densities, which basically accounts to not reaching free flow and, hence, residing still in bound flow. In other words, the systems reaches the fat tail in the loss region.

\subsection{Induced antifragility}

The provided example was chosen to highlight the need for a control algorithm capable of pushing the closed-loop system trajectories to the antifragile regions under the effect of the traffic signal timing. Here we extend the fragility--robustness--antifragility detection with, what \cite{taleb2013mathematical} termed degrees of fragility (i.e. inherited and intrinsic fragility), induced antifragility. Realizing induced antifragility assumes the design of a closed-loop control system which can judiciously compute traffic signals that compensate uncertainty.

As we see in our simple analysis in Figure~\ref{fig4}, intrinsic and inherent fragility characterize the open-loop traffic dynamics, where the second-order effects (i.e. the free flow, bound flow, and congestion) of the MFD can be mapped to the fragile--robust--antifragile spectrum Figure~\ref{fig4} c. Indeed, without a traffic light signal, the system's response to uncertainty is solely described by the shape of the probability distribution of the system's variables and its sensitivity to uncertainty (i.e. intrinsic fragility). Additionally, the space--time--intensity space of uncertainty dictates what is defined as inherited fragility and considers the combined dynamics of system and uncertainty (i.e. transitions between bound flow and free flow and bound flow to congestion, respectively).

\begin{figure} 
\centering
\includegraphics[scale=0.3]{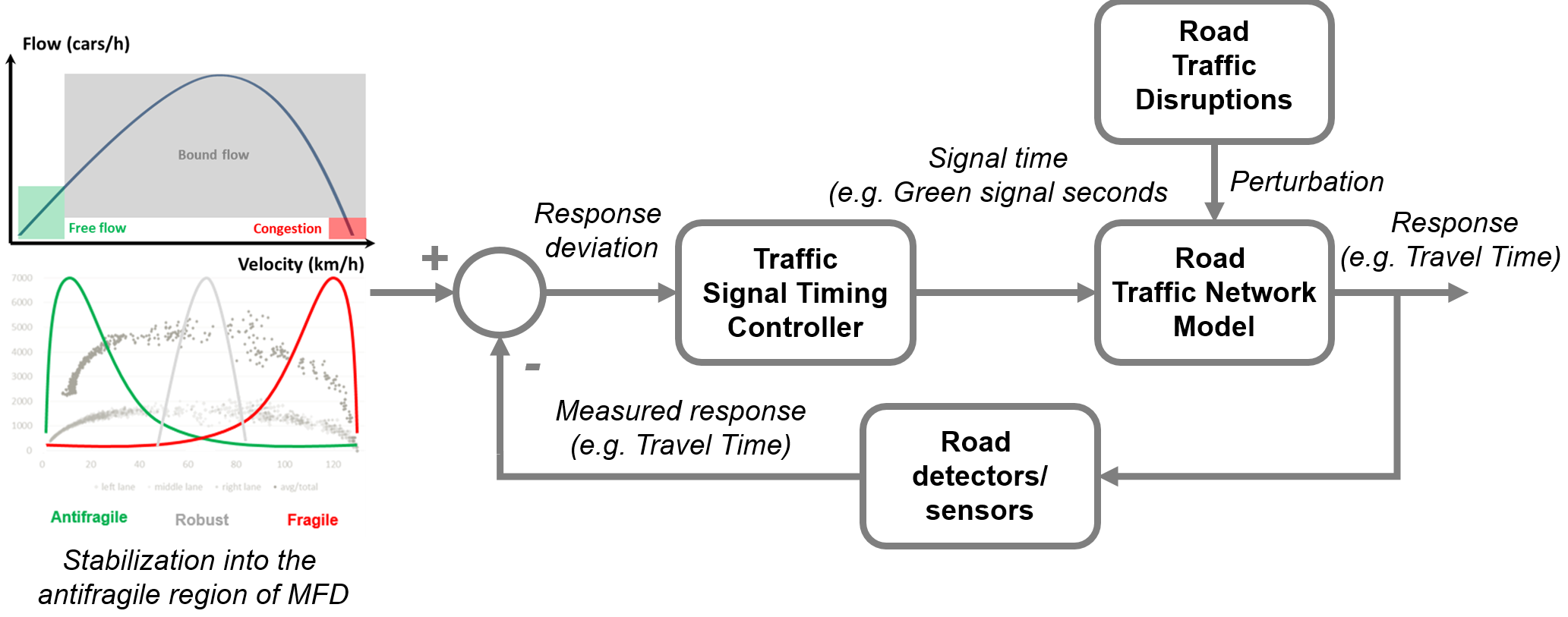}
\caption{Antifragile control closed--loop system. The core idea is to stabilize the road traffic network model in the antifragile region of the MFD so that in the presence of road traffic disruptions, the control law (i.e. signal time) can drive the system to a response (e.g. travel time) which lies in the antifragile region of the MFD, or synonymous, within the bound or free flow regions.}
\label{fig5}
\end{figure}

Our study focuses on designing a control system that pushes the closed-loop traffic dynamics to the antifragile regions of the MFD, which is depicted in Figure~\ref{fig5}. 

As depicted in the closed-loop control diagram, the main problem is, in other words, how to compute a traffic timing signal that pushes the closed-loop behavior to the antifragile region depicted in Figure~\ref{fig4} c. This assumes that the choice of traffic timing signal is robust to uncertainty and can compensate for uncertainty effects, described by positive or negative exposure to volatility for some points of the distribution (see Figure~\ref{fig4} c).

\subsection{Possible paths to induced antifragility}

In order to anchor the reader in the design steps of an induced antifragility control system, we describe how can one modify existing control loops to gain from second-order effects. We will start from a robust system design of \cite{keyvan2012exploiting} and extend it with additional elements capturing those relevant second order effects relevant for gaining under uncertainty: from gain factors of convex-concave characteristics of traffic from the work of \cite{ambuhl2020functional} up to nonlinear effects and generalized fundamental diagram from the work of \cite{ranjan2019emergence} and \cite{knoop2015traffic}. 

\begin{figure} [h!] 
\centering
\includegraphics[scale=0.31]{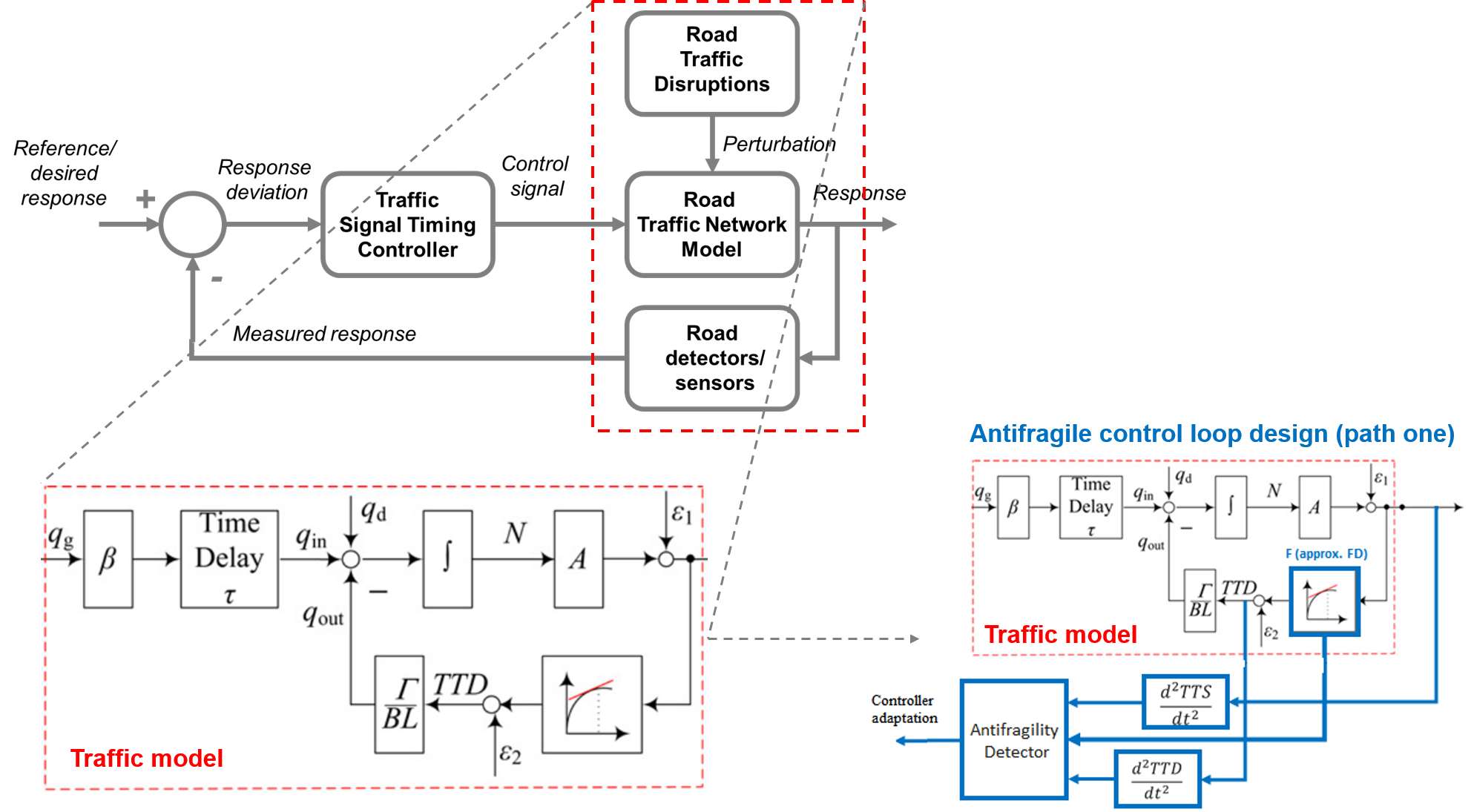}
\caption{Path one to induced antifragile closed--loop control systems: extending robust control with second order effects of the evolving dynamics of a protected region flows. Based on the traffic model used in \cite{keyvan2012exploiting} for robust control design, The Total Time Spent (TTS) reference is mapped by the controller to a gating flow of cars $q_g$ which enters the road network region given a $\beta$ weighted time delay $\tau$ under the impact of disturbances $q_d$ and the output flow $q_{out}$. The TTS is then captured by the evolution of the total number of cars $N$ within the region, whereas the Total Traveled Distance (TTD) is computed using the empirical fundamental diagram $F$ estimated from the data and which describes the $\frac{\Gamma}{BL}$ weighted output flow $q_{out}$.}
\label{path1}
\end{figure}

In order to avoid congestion-caused degradation (i.e. a Total Traveled Distance (TTD) decrease), the critical value (i.e. the value of Total Time Spent (TTS) at which the maximum TTD is attained) in the fundamental diagram is considered as the set value for the robust controller designed by \cite{keyvan2012exploiting}. The control goal was to keep the traffic state of the region around the set value, so that TTD is maximized and the network does not enter the over-saturation area in the fundamental diagram. In order to drive the system towards the desired region of the fundamental diagram mapped to fragile-robust-antifragile continuum, the antifragile controller is updated based on the curvature of both TTS and TTD (i.e. second order effect) and the learnt F (i.e. learnt approximation of the fundamental diagram form observations). The antifragility detector applies some heuristics (i.e. statistical) to update the parameters of the controller. Figure~\ref{path1} captures the design elements to turn the robust control of \cite{keyvan2012exploiting} into an antifragile one.

\begin{figure}[h!] 
\centering
\includegraphics[scale=0.31]{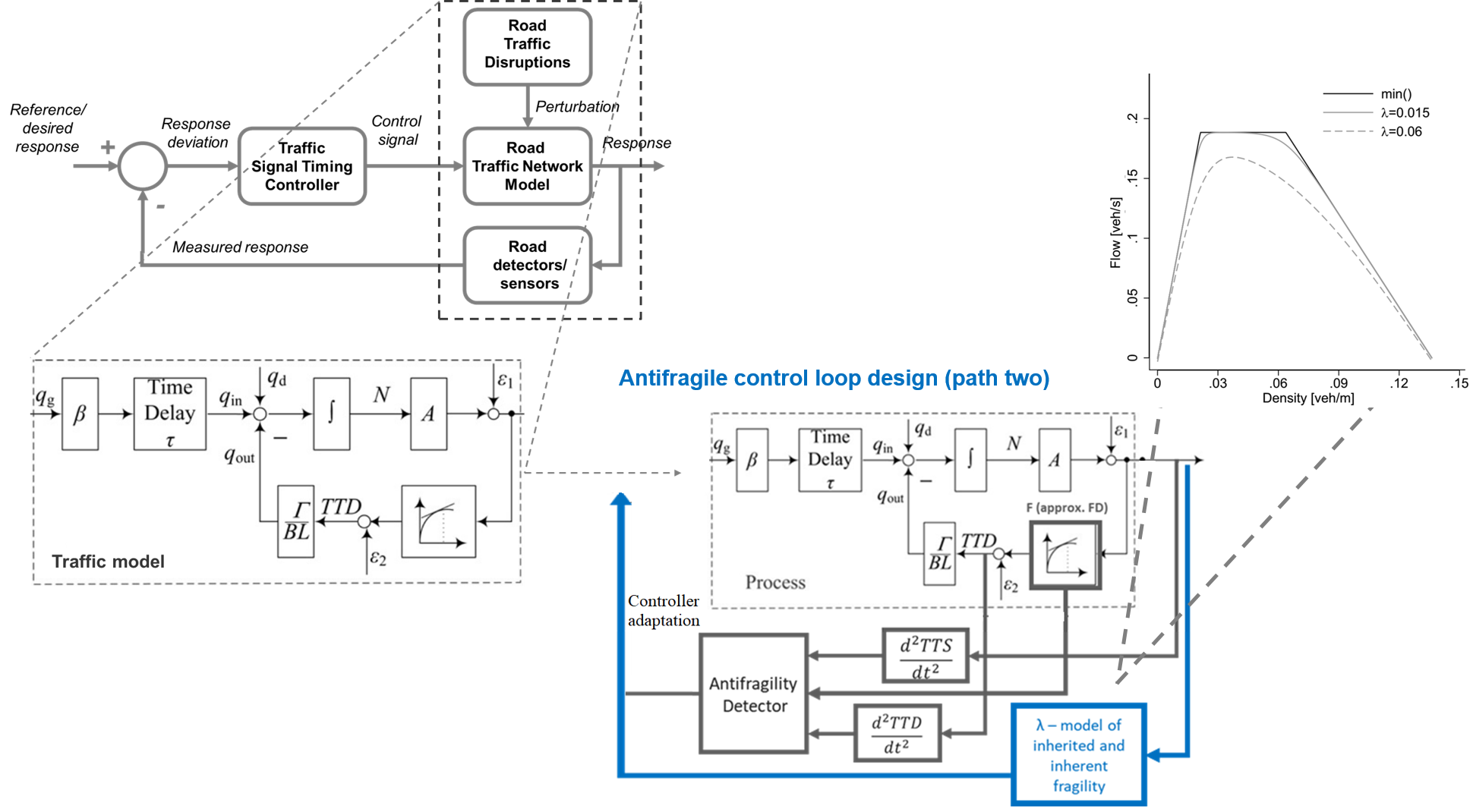}
\caption{Path two to induced antifragile closed--loop control systems: extending robust control with second order effects of the evolving dynamics of the macroscopic fundamental diagram. Considering the traffic model of \cite{keyvan2012exploiting} used for robust control design, the parameter $\lambda$ is a collective and network-wide quantification of flow reducing factors, caused by infrastructure, between-vehicle interactions, and other means of transportation such as cycling and walking, as shown in the study of \cite{ambuhl2020functional}. For the same fundamental diagram, flow values decrease with increasing values of $\lambda$. Smaller values of $\lambda$ indicate that the infrastructure is used more efficiently and that performance losses due to vehicle interactions are smaller, as demonstrated in the study of \cite{ambuhl2020functional}.}
\label{path2}
\end{figure}

Going beyond the "traditional" control loop of \cite{keyvan2012exploiting}, the work of \cite{ambuhl2020functional} introduced a new parameter ($\lambda$) that can be useful in fundamental diagram applications from analyzing urban congestion to traffic control. In this functional form, the fundamental diagram can be either estimated from measurements or defined a-priori, either analytically or with additional measurements in the network, while the smoothing is quantified with a single parameter $\lambda$. In order to drive the system towards the desired region of the fundamental diagram mapped to fragile-robust-antifragile continuum, the antifragile controller uses $\lambda$.  The parameter $\lambda$ becomes then a collective and network-wide quantification of flow reducing factors, caused by infrastructure, between-vehicle interactions, and other means of transportation such as cycling and walking. For the same fundamental diagram, flow values decrease with increasing values of $\lambda$. Smaller values of $\lambda$ indicate that the infrastructure is used more efficiently and that performance losses due to vehicle interactions are smaller. This behavior is captured in the Figure~\ref{path2} where the $\lambda$ gain, introduced by \cite{ambuhl2020functional}, is used in the controller adaptation scheme after the antifragility detection on the second order effects of the dynamics.

\begin{figure} [h!] 
\centering
\includegraphics[scale=0.32]{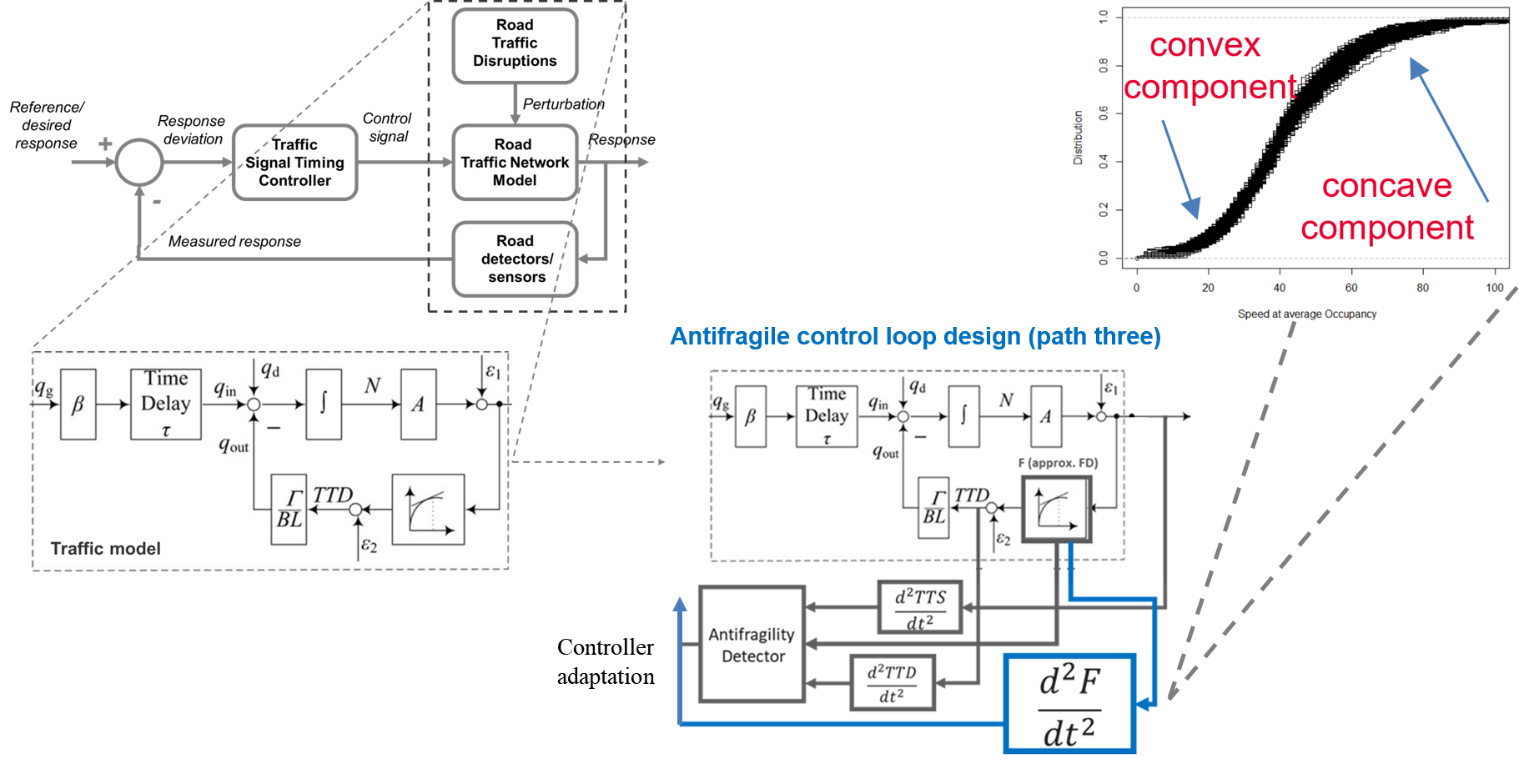}
\caption{Path three to induced antifragile closed--loop control systems: extending robust control with second order effects of the evolving dynamics given by the shape of the traffic fundamental diagrams. Considering the traffic model of \cite{keyvan2012exploiting} used for robust control design, the study of \cite{ranjan2019emergence} is a good reference point to start. The stable relationship between the space-averages of speed, flow and occupancy are insufficient to infer a robust relationship and the emerging fundamental diagram cannot be guaranteed to be stable if traffic interventions. Second order effects (i.e. concave or concave shape of the distribution) in responses (i.e. speed distributions over average occupancy) can be used to adapt controller operation.}
\label{path3}
\end{figure}

The study of \cite{ranjan2019emergence} has demonstrated that observing the stable relationship between the space-averages of speed, flow and occupancy are not sufficient to infer a robust relationship and the emerging fundamental diagram cannot be guaranteed to be stable if traffic interventions (i.e. traffic light control to react to disruptions) are implemented. Second order effects in responses (i.e. speed distributions) can be used to adapt controller operation. What does this imply for the use of an fundamental diagram for traffic control? Metering would affect the observed shape of the fundamental diagram and that it therefore does not predict the effect of metering. So it should now be clear that merely observing a stable relationship between space-averages of speed or flow versus occupancy is not sufficient to infer the existence of a robust relationship that can be used for traffic control. Quantifying the curvature of individual responses to signal re-computation/metering, as suggested by \cite{ranjan2019emergence}, could enable a correction of the fundamental diagram approximation in $F$ as shown in Figure~\ref{path3}.

\begin{figure} [h!] 
\centering
\includegraphics[scale=0.31]{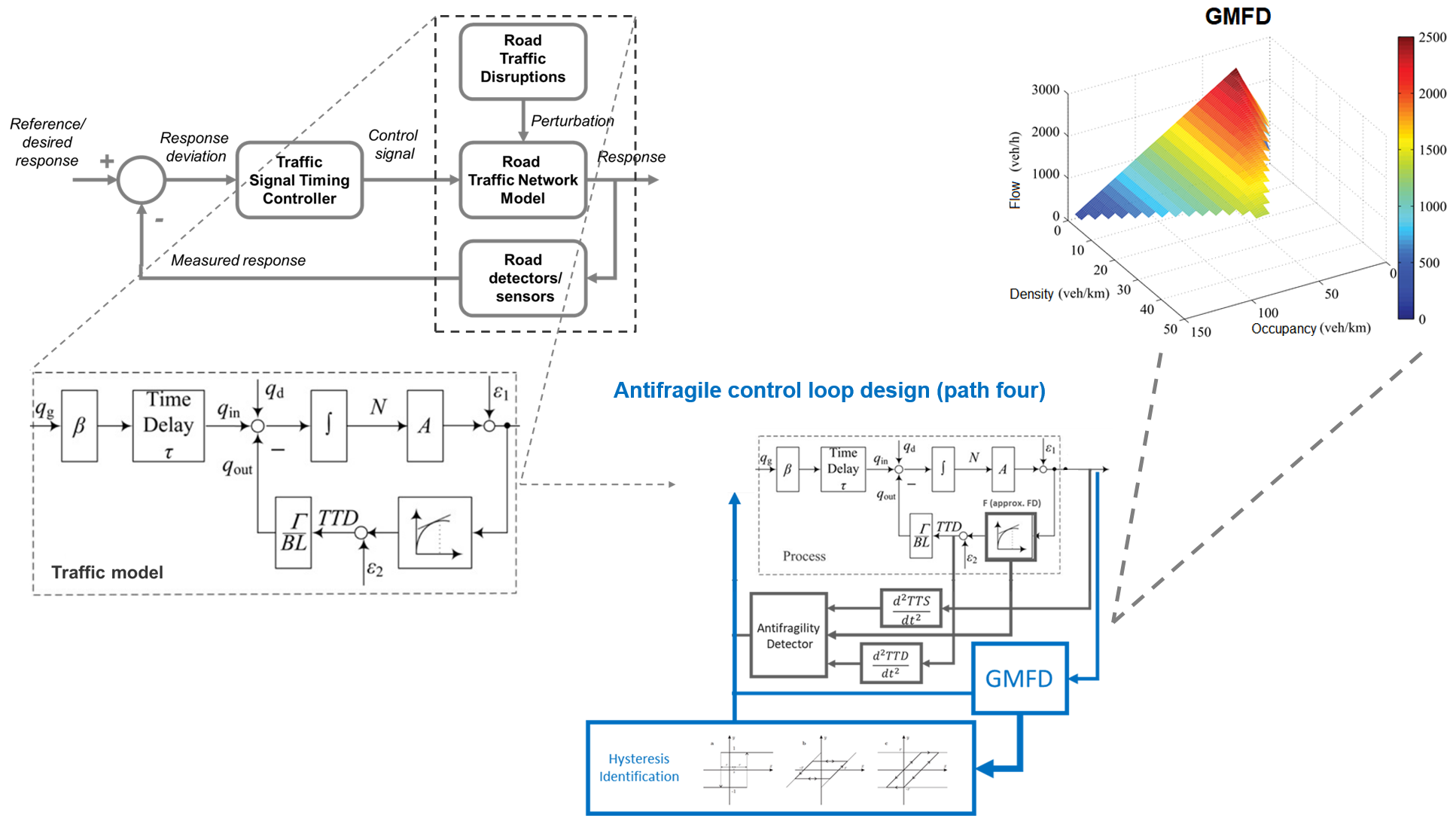}
\caption{Path four to induced antifragile closed--loop control systems: extending robust control with second order effects of the evolving dynamics and their in-homogeneity and nonlinear shape. Considering the traffic model of \cite{keyvan2012exploiting} used for robust control design, \cite{knoop2015traffic} explored the fact that production is a continuous function of accumulation and spatial in-homogeneity of density and learn the production function ((i.e., the average flow of vehicles per unit of time)). The study of \cite{knoop2015traffic} proposed the use of Generalized Macroscopic Fundamental Diagrams (GMFD). This can be used to detect relevant nonlinear effects in the response and extend the closed-loop control loop to attain antifragile responses.}
\label{path4}
\end{figure}

Finally, the work of \cite{knoop2015traffic} has shown that Macroscopic Fundamental Diagram (MFD) when aggregated over an area, describes the relationship between accumulation and production and is quite crisp. This can be an aspect of the law of large numbers: the more data aggregated, the smaller the influence the differences in drivers characteristics have. For control purposes, it is very useful to have a strict relationship on the basis of which control can be applied. Traffic dynamics in congested networks inherently leads to traffic in-homogeneity, because congestion, under increasing traffic volume loaded onto the network at some point, will set in at one of the (potentially many) bottlenecks in a network. But, exploiting the fact that production is a continuous function of accumulation and spatial in-homogeneity of density, a solution proposed by \cite{knoop2015traffic} is to extract the generalized MFD or learn the production function (i.e., the average flow of vehicles per unit of time). Then, in the antifragile design we can capture the hysteresis effects found in the MFD using the generalized MFD of \cite{knoop2015traffic} to employ in the antifragile control scheme, described in Figure~\ref{path4}. For instance, it can be used for estimating speed in a (sub-) network, and traffic can be guided over the faster routes.

Instead of following this extension approach, we will focus, in the rest of the manuscript, on the oscillator-based modelling and control. We will expand the "recipe" to design the induced antifragility closed-loop control system and its evaluation.

\subsection{Contributions}

The main contribution of this study fundamentally revolves around another instantiation of the novel framework of antifragile control for road traffic optimization. We consider the problem of induced antifragility of an oscillator-based road network traffic model by means of stabilizing the closed-loop system in the detected antifragile regions of the macroscopic fundamental diagrams. In other words, we design a system that judiciously computes signal timings for traffic lights in spatially connected intersections such that measures of performance (e.g. travel time) are optimized.

As designing and building such a closed-loop system is a "tour-de-force", we identify the main components of our contribution and highlight for our readers what are the highlights expanded in the following sections.

\begin{itemize}
\item we systematically characterize road traffic disruptions in a space--time--intensity reference system
\item we design a mapping from the traffic equilibrium regions in an MFD to the fragile--robust--antifragile continuum
\item we employ an oscillator-based model for the road network traffic model that captures both temporal effects (e.g. load (flow) interactions in traffic dynamics under disruptions) and spatial structure (i.e. coupled streets with different capacities and geometries)
\item we formalize the closed-loop control problem as a stabilization on MFD and investigate antifragile control synthesis mathematical underpinnings for stabilization
\item we design and implement an antifragile controller capable of handling external uncertainty by demonstrating closed-loop systems benefit from variability and are less sensitive to volatility (i.e. measure of scale of a distribution in Figure~\ref{fig4})
\item we evaluate and discuss our results from this new instantiation of antifragile control on a real-world road traffic dataset which contains 59 days of real urban road traffic data from 8 crosses in a city in China
\end{itemize}

\section{Materials and Methods}

In this section, we introduce the models and tools we employed in our study. We start with the oscillator-based network model of a large urban road traffic scenario in China based on the work of \cite{axenie2021obelisc} and the data from \cite{cristian_axenie_2021_5025264} available at \url{https://doi.org/10.5281/zenodo.5025264}. This large network traffic model will be used both in open-loop (no feedback controller) and closed-loop control configurations. Afterwards, we formally introduce the antifragile control framework realization for the road traffic scenario. This formalism consolidates and extends the mathematical apparatus and controller synthesis from the work of \cite{axenie2022antifragile} and the techniques suggested by \cite{taleb2013mathematical}. Finally, we introduce the formalism and parametrization of alternative (and relevant) road traffic control and optimization methods (i.e. Optimal Control (using Mixed Integer Linear Programming (MILP)) and Robust Control) to comparatively evaluate and probe the benefits of antifragile control.

\subsection{Oscillator-based road traffic model}

Both road traffic and road traffic control signals have a strong periodic behavior. On a fast timescales, the periodicity of traffic light signals determines fluctuations in the road traffic flow that reflect in daily periodic patterns (i.e. slow timescales). Take for instance the "camel-back" profile of traffic over working week days, which peaks around 9 AM and 6 PM. Such a perspective motivates us to describe traffic light phasing phenomenon as a repeated synchronization problem, in which a large network of oscillators, each representing a traffic light controlling a possible movement direction in a cross, spontaneously locks to a common operation phase. This amounts to a state in which opposite directions will have a green light and adjacent directions will have no right of pass. 
Under these assumptions, the phase duration adjustment factor (i.e. the green time) is computed as a function of the oscillator time to synchronization. Such a modelling approach has also been considered in the work of \cite{chedjou2018review}, \cite{nishikawa2004dynamics}, \cite{fang2013matsuoka} and, of course, \cite{axenie2021obelisc}. The underlying intuition is the following: 1) each of the traffic light oscillators is injected with (sensors' measured) traffic flow data impacting its local dynamics (i.e. local computation of green time given neighboring oscillators), and 2) the oscillator network (i.e. mapped on the traffic lights of an urban region) converges to a steady state used to extract adaptive factor to adjust the traffic light phases. Despite the inevitable differences in the natural oscillation frequencies and injected traffic flow data of each oscillator, the network ensures that each of the coupled oscillators repeatedly locks phase (i.e. synchronizes given the interactions with the neighboring oscillators within a cross and between crosses). 
The model we employ in this study is based on the model of \cite{axenie2021obelisc} which extends the basic Kuramoto oscillator in \cite{strogatz2000kuramoto} as described in Equation~\ref{oscillator_eq}.
\begin{equation}
\frac{d\theta_i(t)}{dt} = \omega_i(t) + k_i(t)\sum_{j=1}^N{A_{ij}sin(\theta_j(t) - \theta_i(t))} + F_i sin(\theta^*(t) - \theta_i(t))
\label{oscillator_eq}
\end{equation}
where:\\
$\theta_i$ - the amount of green time of traffic light $i$ \\
$\omega_i$ - the frequency of traffic light $i$ oscillator \\
$k_i$ - the flow of cars passing through the direction controlled by oscillator $i$ \\
$A_{ij}$ - the static spatial adjacency coupling between oscillator $i$ and oscillator $j$\\
$F_i$ - the coupling of external perturbations (e.g. maximum cycle time per phase)\\
$\theta^*$ - the external perturbation (e.g. traffic signal limits imposed by law)\\

The model assumes that the change in allocated green time $\theta_i$ for a certain traffic light $i$, for a certain direction, depends on the: 1) the internal frequency of the corresponding (traffic light) oscillator $\omega$; 2) the current flow of cars $k_i$ in that direction; 3) the spatial coupling $A_{ij}$ of the oscillators through the street network that weights the impact of a nonlinear periodic coupling of the oscillators $sin(\theta_j(t) - \theta_i(t))$; and 4) the external perturbation $\theta^*$ with weight $F_i$ which ensures, for instance, that the output of the system stays in the bounds of realistic green time values imposed from the traffic laws.
Given the known topological layout of the road network and the computed green times of each of the oscillators, when the dynamics converge (i.e. the solution of the differential equation \ref{oscillator_eq}), we infer the actual adaptive factor (i.e. delta green time) to be applied to the traffic light phase duration between adjacent (coupled) oscillators corresponding to adjacent moving directions. More precisely, given the steady state value of the green time (i.e. the solution $\theta_i(t_f)$), we calculate the phase duration as the time to synchronization of each oscillator relative to the ones coupled to it. From the dynamics synchronization matrix $\rho$ at each time $t$ the phase duration update is calculated as $\underset{t}{\arg \max} \{ \rho(t) > \tau \}$ where $\rho_{ij}(t) = cos(\theta_i(t) - \theta_j(t))$ and $0 < \tau < 1$. 

\begin{figure}[!h]
\centering
\includegraphics [scale=0.3] {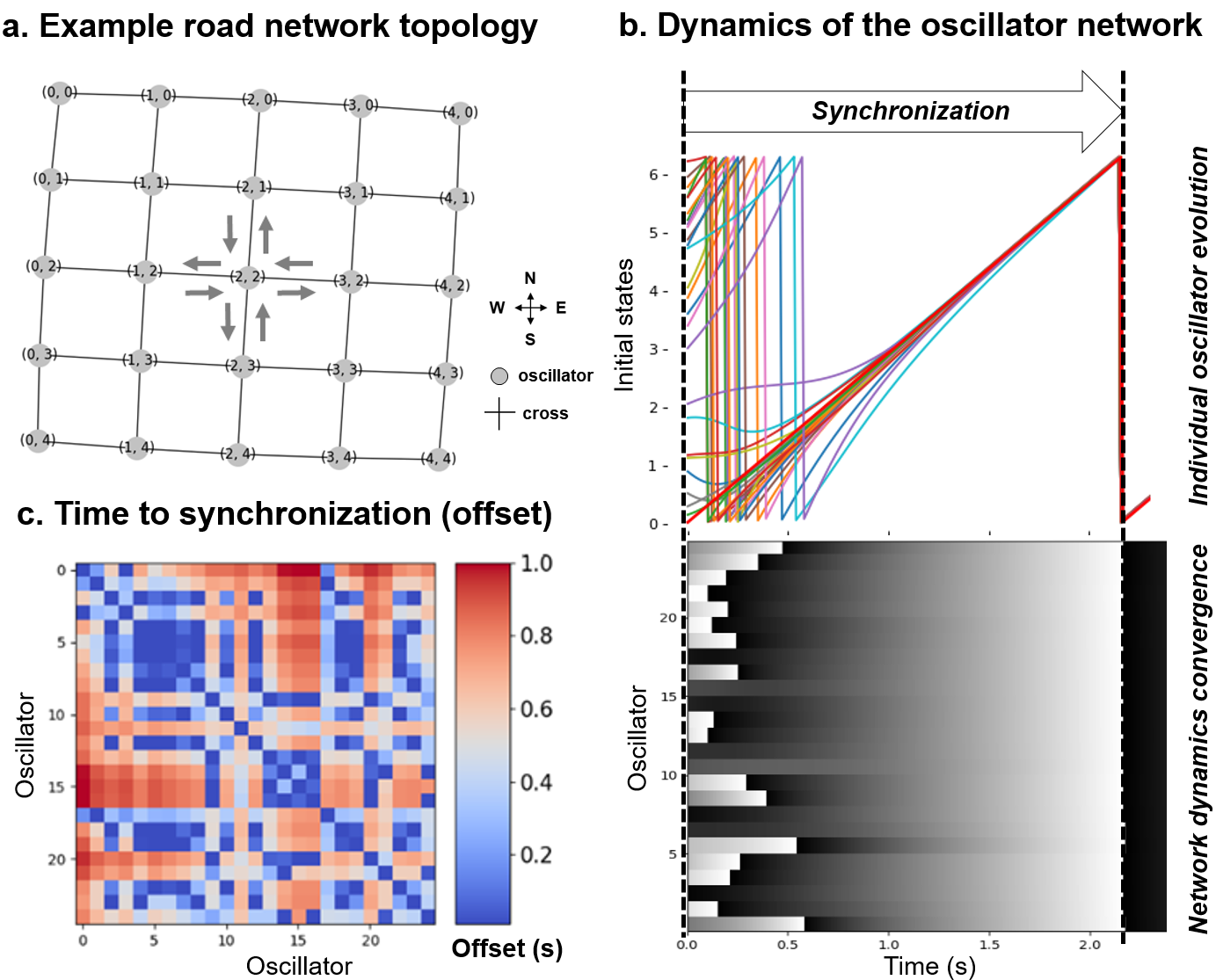}
\caption{Oscillator-based network model dynamics. a) Network geometry and spatial coupling of oscillators. b) Dynamics of the oscillator network model from initial conditions and without control signals, phase lock under "free evolution" after a transient. c) Time to synchronization of one oscillator w.r.t. spatially connected neighbors as a quantification of offset of green time to compute the green time phase.} 
\label{fig6}
\end{figure}
In order to ground the analytic formulation, we describe a simple, regular $5\times5$ lattice composed of $N=25$ oscillators describing an ideal road network. For simplicity, in this example, each oscillator is responsible for an entire cross (i.e. the 4 adjacent directions: N, S, W, E) and the spatial coupling $A_{ij}$ is given by the topology of the lattice, as shown in Figure~\ref{fig6} a. Here, for each oscillator $i$ its dynamics is described by the superposition of its natural oscillation frequency $\omega_i$ and the cumulative impact of neighboring (coupled through $A_{ij}$) oscillators weighted by the flow of cars $k_i$ through the cross controlled by oscillator $i$. Figure~\ref{fig6} b describes the internal dynamics of such a network model for traffic control where given the different initial conditions of each oscillator, the coupling dynamics enforces consensus after some time (i.e. $\approx 2.1 s$). The steady state is then used to extract the actual phase duration by simply calculating the time to synchronization $\underset{t}{\arg \max} \{ \rho(t) > \tau \}$, as a per oscillator relative time difference, from the $\rho$ matrix in Figure~\ref{fig6} c. Here, the choice of $\tau$ determines how fast a suitable steady state is reached.

\subsection{Antifragile control}

As coined in the book of \cite{taleb2012antifragile}, antifragility is the property of a system to gain from the uncertainty, randomness, and volatility describing its operating environment and the system's exposure to events. This is opposite to what fragility would incur but it also goes beyond robustness. An antifragile system's response to perturbations and stress is typically beyond robust, in that exposure to small amplitude (high-frequency) stressors can strengthen the future response of the system to high amplitude events through a strong anticipation component.
In the seminal work of \cite{taleb2013mathematical}, along the mathematical formulation of the spectrum fragile -- robust -- antifragile, Taleb introduced a taxonomy of fragility. 

This is the departing point of our development. In the original classification of \cite{taleb2013mathematical}, fragility could be either local (i.e. characterizing the system model and its parameter distributions) or inherited (i.e. determined by the system's environment or exogenous signals). Additionally, using the so called "transfer function" one could map a system's sensitivity to stress to its response shape curvature (i.e. the second order effect). The goal of our control theoretic approach is to "push" a system's behavior from the fragile or robust regions of the spectrum to the antifragile region. In other words, we want to induce antifragility! Of course, considering the inherent (and immutable) local and inherited fragilities of the system, we aim to reach induced antifragility by means of a properly chosen control law that steers the system trajectories in the desired region.
When looking at control systems, inducing such a behavior to a feedback control loop accounts for a novel design and synthesis approach, originally introduced in \cite{axenie2022antifragile}. The main principles behind this approach and, fundamentally, the induced antifragility are: (1) redundant overcompensation that makes the system trajectory overshoot (in a controlled way) in order to build extra-capacity in anticipation; (2) structure-variability, achieved through a discontinuous control law that can induce (high-frequency) stress while still driving the system's trajectory in the desired region; and 3) convexity of the system response, once reached the desired region, under the effect of the control law.

\subsubsection{Formalism}

In our previous study, that laid the foundations of the control-theoretic approach to induced antifragility, we cast the control design in the geometric control and Riemannian geometry objects, formally described in \cite{lee2006riemannian}. This enabled us to work in a coordinate-free space where we relied on the embedding of a manifold into a larger dynamical space that allowed for simpler control law definitions suitable for manifolds with curvature (i.e. second order effects). In the current study, we consolidate this framework and relax some assumptions we previously made. 

\textbf{Preliminaries}

In our mathematical development of the study, we will always consider the most general form of the dynamical system in Equation~\ref{oscillator_eq}, namely
\begin{align}
\dot{\theta_i}  = f_i(t, \theta_i, u_i)
\label{eq1}
\end{align}
where the state vector $\theta_i$ takes values on a smooth manifold $\Theta$, $t$ is the time, $u_i$ is the control law for oscillator $i$, and $f_i$ is a smooth nonlinear function.
In our initial realization of antifragile control in \cite{axenie2022antifragile}, we framed our problem as a tracking control problem where a reference (i.e. desired) vector $\theta_{id}$ (i.e. the desired dynamics $\theta_i=\theta_{id}$) was used to compute an error function $\varphi(\theta_i, \theta_{id})$ subsequently used in the variable structure control design, as described in \cite{utkin2013sliding} and \cite{slotine1991applied}. More precisely, the design objective was the synthesis of a control law $u$ such that the system trajectories reach and stay on a manifold $\sigma$, in other words $\sigma(t, \theta_i, \theta_{id}) = 0$, for which 
\begin{align}
\varphi(\theta_i, \theta_{id}) = \theta_i - \theta_{id} &\\
\sigma(\theta_i, t) = \left(\frac{d}{dt} + \lambda\right)\varphi(\theta_i, \theta_{id}), \lambda > 0
\label{eq2}
\end{align}
This allowed us to deal with uncertainty by reaching and staying under the effect of $u_i$ on $\sigma$, which now describes the overall system dynamics. In other words, given that $\theta_{id}(0) = \theta_i(0)$, our control problem of tracking $\theta_i=\theta_{id}$ is equivalent to that of remaining on $\sigma(t, \theta_i)$ for all $t > 0$.
Opposite to our initial realization of antifragile control, we replace the tracking problem $\theta_i=\theta_{id}$, or in other words $\lim_{t \to \infty}\varphi(\theta_i, \theta_{id}) = 0$, with a stabilization problem in $\sigma$. We compute the manifold $\sigma$ as a discontinuous element (i.e. variable structure assumption still holds) and its second derivative $\ddot{\sigma}$ as a function of the rate of change of the control law $u$. This allows us to drop the dependency on  $\varphi(\theta_i, \theta_{id})$ and many other benefits, as we will see in the following section.
Second, the redundant overcompensation feature of antifragile control, is now generalized through the use of a second order sliding mode controller and not through a manifold Proportional Derivative (PD) control and a (problem-dependent) choice of a Riemannian transport map (i.e. describing the actual dynamics transformation that pushes the system towards the desired manifold along a geodesic). Using second order sliding modes allows us to reduce the chattering effects (i.e. excessive switching while approaching the $\sigma$) by using a well-behaved dynamics given by the second-order differentiation of the sliding manifold $\ddot{\sigma}$ that depends on the rate of change of the synthesized control law $\dot{u}$.
The above generalizations over the previous realization of antifragile control are motivated on one side by the type of system we try to control (i.e. network of coupled oscillators maximising road traffic flow and minimizing travel time) and, second, by the flexibility and accuracy second-order sliding modes can bring in nonlinear control (as introduced by the seminal work of \cite{levant1993sliding}, \cite{levant2005homogeneity}, and, of course, the very good survey of \cite{bartolini2003survey}).

\textbf{Formulating a Stabilization Problem}

In order to enable the desired motion of the closed-loop system, we consider a stabilization problem of the system's dynamics $\dot{\theta_i}$, on a known manifold (i.e. sliding surface) $\sigma$. Yet, in order to ensure motion accuracy (i.e. sliding mode) and avoid high-frequency control actions detrimental in traffic control, we consider high-order motion manifolds (i.e. sliding surfaces) and synthesize a sliding mode controller as suggested in \cite{slotine1991applied} and \cite{utkin2013sliding}. This type of control is a very good candidate for practically reaching closed-loop antifragility and the preliminary work of \cite{axenie2022antifragile} confirms this. We will motivate this in the following the theoretical underpinnings of the control synthesis. Note that for the following we replace $\theta_i$ with $\theta$ to simplify notation while still keeping the meaning of the single oscillator dynamics within the network.

As we previously motivated, we convert the $\theta_{id}$ reference (i.e. desired dynamics) tracking problem in a stabilization problem in $\sigma$. This appears immediately if we differentiate $\sigma$ once in Equation~\ref{eq2} for the control law $u_i$ to appear (i.e. simply differentiate Equation~\ref{eq2} and replace $\dot{\theta_i}$ with Equation~\ref{eq1}). This is also intuitive, as we do not have any $\theta_d$ available, but only the sum of all allocated green time in a cross/region. Stabilizing to $\sigma$ accounts in this case to drive the network of oscillators to a phase lock in a certain amount of time (i.e. reaching time or synchronization time). Furthermore, bounds on the sliding manifold $\sigma$ can be directly translated into bounds on the tracking error vector $\varphi(\theta_i, \theta_{id})$, and therefore $\sigma$ represents a true measure of tracking performance as demonstrated in \cite{slotine1991applied}. 

We are interested in the high-order effects of the sliding surface $\sigma$ in order to not only determine the motion of the closed-loop system (i.e. phase lock on an optimal configuration of traffic light green time allocations) but also to make judicious decisions on the control law signal $u$ (i.e. total green time reaches police imposed values). Let's consider the $l$--th derivative of $\sigma$ as being 
\begin{align}
\sigma^{(l)}(\theta, t) = (\frac{d}{dt})^{(l)}\sigma(t, \theta(t, \epsilon))
\label{eq3}
\end{align}
a uniformly bounded function of $\epsilon$. Then for the steady state part of $\theta(t,\epsilon)$ there exist positive constants $C_1, C_2, ..., C_{l-1}$ such that the following equations hold
\begin{align}
\lVert \dot{\sigma} \rVert \leq C_1\tau^{l-1} \\
\lVert \ddot{\sigma} \rVert \leq C_2\tau^{l-2} \\
\vdots \\
\lVert \sigma^{(l-1)} \rVert \leq C_{l-1}\tau
\label{eq4}
\end{align}
if $\tau(\epsilon) > 0$ is the smallest time interval of smoothness of the piece-wise smooth function $\theta(t,\epsilon)$. This demonstrates that considering high-order sliding manifolds allows for a smoothed control signal. This will turn out useful in out traffic control problem where the choice of green time needs to judiciously follow a smooth pattern to allow connected oscillators to lock phase.

If we consider Equation~\ref{eq1} in Filippov's sense (see work of \cite{blagodatskikh1985differential} for formalism), we can then replace it by an equivalent differential inclusion $ \dot{\theta} \in F(\theta, t, u)$ where if the vector field $F$ is continuous then the set value $F(\theta_0, t_0, u_0)$ is the convex closure, as demonstrated in the work of \cite{filippov1963differential}. We now integrate this new formulation in the framework initially introduced in \cite{axenie2022antifragile} where the Riemannian geometry offers the mathematical framework to develop the control synthesis. 

Let $\Gamma$ be a smooth manifold. The set $\Gamma$ is called first order sliding point set. The second-order sliding point set is the set of points $\theta\in\Gamma$, where $F(\theta)$ lies in the tangential space $T_\theta\Gamma$ to the manifold $\Gamma$ at the point $\theta$. We remind the reader that $T_\theta$ is the parallel transport map that describes the actual dynamics transformation pushing the system towards the desired manifold along a geodesic. Then there exists a second-order sliding mode on $\Gamma$ in the vicinity of a second-order sliding point $\theta_0$ if in this vicinity of the point $\theta_0$ is an integral set (i.e. it consists solutions in the Filippov sense -- see \cite{filippov1963differential}).

In this case, all possible velocities $F$ lie in the tangential space $T_\theta(F)$ and even when a switching error is present, the state trajectory is tangential to the manifold $\Gamma$ at the time of leaving. Back to the formulation in Equation~\ref{eq1}, we can now extend the description of the closed-loop as following
\begin{align}
\dot{\theta} = f(t, \theta, u) &\\
u = U(t, \theta, u) &\\
\xi = \Psi(t, \theta, \xi)
\label{eq5}
\end{align}
where $U$ now describes the first-order sliding (i.e. recall that differentiating $\sigma$ one recovers $u$) and $\xi$ now describes the second-order sliding mode, basically the velocity of the control law $u$. Importantly, in second-order sliding mode $U$ is a continuous function and $\Psi$ is bounded discontinuous, hence the formulation can be written as a continuous control
\begin{align}
\dot{\theta} = f(t, \theta, u_{eq}(t, \theta)),&\\
\label{eq6}
\end{align}
where $u_{eq}$ is the equivalent control law evaluated from 
\begin{align}
\dot{\sigma}(t, \theta) = \frac{\partial}{\partial t}(\sigma(t, \theta)) + \frac{\partial}{\partial \theta}(\sigma(t, \theta))f(t, \theta, u_{eq}) = 0
\label{eq7}
\end{align}
or 
\begin{align}
\dot{\sigma}(t, \theta, u) = L_u\sigma(t, \theta)
\label{eq8}
\end{align}
where
\begin{align}
L_u(\cdot) = \frac{\partial}{\partial t}(\cdot) + \frac{\partial}{\partial \theta}(\cdot)f(t, \theta, u)
\label{eq9}
\end{align}
is the total Lie derivative with respect to $\dot{\theta} = f(t, \theta, u)$ when $u$ is constant. Then, we can clearly see that the curvature of the sliding manifold $\sigma$ can be computed by
\begin{align}
\frac{d^2}{dt^2}\sigma = \frac{d}{dt}(L_u\sigma) = L_uL_u\sigma +  \frac{\partial}{\partial u}L_u\sigma\dot{u}
\label{eq10}
\end{align}
If we let $C = L_uL_u\sigma$ and $K = \frac{\partial}{\partial u}L_u\sigma$, then we can write Equation~\ref{eq10} as 
\begin{align}
\ddot{\sigma} = \frac{d^2}{dt^2}\sigma = C + K\dot{u}
\label{eq11}
\end{align}
with $\|C\| \leq C_0$ and $0 < K_m \leq K \leq K_M$.

\begin{figure} 
\centering
\includegraphics[scale=0.45]{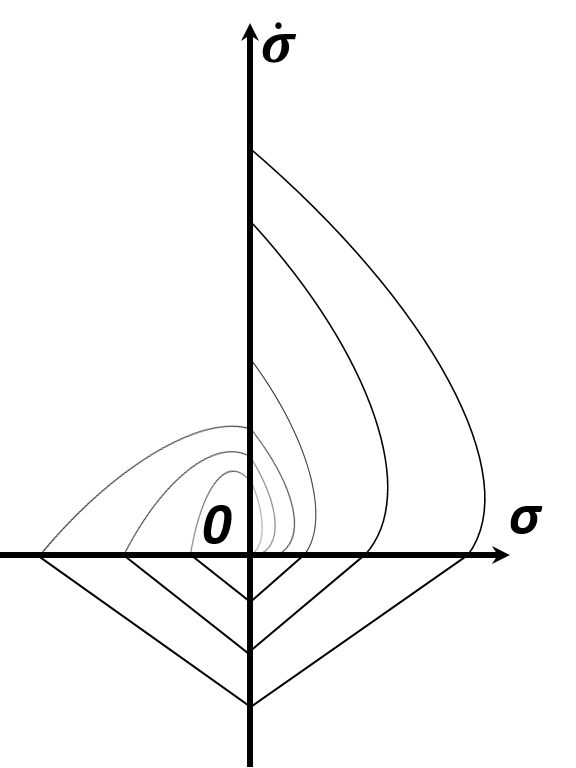}
\caption{System's motion in second-order sliding mode. Sliding manifold curvature depends on the velocity of the control law $\dot{u}$.}
\label{fig7}
\end{figure}

The solution of Equation~\ref{eq11} that keeps the system on the sliding manifold in sliding mode is obtained from $\ddot{u} = 0$ and depicted in Figure~\ref{fig7}. The velocity of the control law $\dot{u}$ is given by
\begin{equation}
 \dot{u} =
    \begin{cases}
      -u & \text{if $|u| > 1 $}\\
      -\alpha sign(\dot{\sigma} - g(\sigma)) & \text{if $|u| \leq 1$}
    \end{cases}   
   \label{sign}    
\end{equation}
where $g = -\beta \sign(\sigma |\sigma|^\gamma), 0.5 \leq \gamma \leq 1$. The positive sub-unitary parameters $\alpha$ and $\beta$ represent the \textit{overcompensation factor} and the \textit{anticipation factor} respectively. Additionally, we can have finer control when the system moves close to the sliding manifold $\sigma$ by considering a boundary $\Phi$ so that the manifold reaching (i.e. sliding manifold attraction) is higher for $\dot{\Phi} < 0$ and lower for $\dot{\Phi} > 0$. Then we can rewrite the control law velocity $\dot{u}$ by replacing the $\sign$ function with a $sat$ function such that $\sign(\sigma)$ can be replaced by $sat(\frac{\sigma}{\Phi})$ and then 
\begin{equation}
 \dot{u} =
    \begin{cases}
      -u & \text{if $|u| > 1 $}\\
      -\alpha sat(\frac{\dot{\sigma}}{\dot{\Phi}} - \beta\frac{\sigma|\sigma|^{\gamma}}{\Phi^2}) & \text{if $|u| \leq 1$}
    \end{cases}   
   \label{sat}    
\end{equation}
These results are consistent with the seminal work and study of \cite{levant1993sliding} and the work of \cite{song2020new} in that the approach guarantees the finite-time convergence to a predefined real sliding mode with controllable overcompensation through the curvature of the sliding manifold, as shown in Figure~\ref{fig7}.

In the following we motivate the need for structure variability to reach closed-loop antifragile behavior. We now particularize the theoretical apparatus for the network of oscillators in Equation~\ref{oscillator_eq}.

\textbf{Structure Variability}

Recall that our goal is to design a sliding manifold to which the controlled closed-loop system trajectories must belong. Sliding mode control, our candidate for the practical realization of antifragile control, is the core of variable structure systems as shown in the work of \cite{utkin1977variable}. Let's now consider the properties of sliding modes in some greater detail. 

First, in order to detect antifragility through the use of sliding mode control, the trajectories of the system's state vector must belong to manifolds of lower dimension than that of the whole state space, therefore the order of differential equations describing sliding motions is also reduced. We previously demonstrated that 1-st order sliding mode control can drive closed-loop nonlinear systems to the antifragile region of their dynamics using a discontinuous control law (see \cite{axenie2022antifragile} for details) using redundant overcompensation and a high-frequency control law. In the current work, we extend the idea to 2-nd order sliding modes and the impact that the curvature of the sliding manifold has upon closed-loop system motion. Most of the work in this area, covered in the excellent review of \cite{bartolini2003survey}, only considers the basic 2-nd order sliding mode control synthesis where the $\alpha$ and $\beta$, namely the \textit{overcompensation factor} and the \textit{anticipation factor} are updated to confine the closed-loop system to the higher derivative of the control variable.
\begin{align}
\ddot{\sigma} = C + K\dot{u},~u(t) = -\alpha(t)Usign(\sigma - \beta\sigma_M) \\
\label{eq12}
\end{align}
where $\beta \in [0,1)$ and, 
\begin{equation}
\alpha(t) =
    \begin{cases}
      1 & \text{if $ (\sigma - \beta\sigma_M)\sigma_M \geq 0  $}\\
      \alpha^* & \text{if $(\sigma - \beta\sigma_M)\sigma_M < 0$} ,\\
    \end{cases}   
   \label{alpha}    
\end{equation}
with $\alpha^* \in [0, \frac{\beta}{10})$ and $\sigma_M$ is the last extremal value of $\sigma$, typically $\sigma_M = \sigma(t_i)$, where $t_i$ is the initial time instant.

Second, in most of practical systems the sliding motion is control-independent and, is determined merely by the properties of the control plant and the position (or equations) of the discontinuity surfaces. This enables the closed-loop system under sliding motion to reach the antifragile behavior. Furthermore, the initial problem can be decoupled into separate lower dimension sub-problems in which the control is "spent" simply on constructing a sliding mode and the requisite character of motion across the intersection of discontinuity manifolds is given by an acceptable choice of their equations. 
For the antifragile control synthesis we consider higher-order sliding modes, in order to obtain smoother dynamics in the vicinity of the sliding manifold. This assumes that, if the order is higher an additional constraint is typically built, a linear combination of the original constraint (i.e. sliding manifold) and its successive total time derivatives, as demonstrated in the work of \cite{levant2005homogeneity}. In this formulation, the successive derivatives of $\sigma$ do not depend on the control $u$, and $\sigma^{(l)}$ depends linearly on $u$ or its velocity with non-zero coefficients $L_a^{(l-1)}L_b\sigma$ given that the system dynamics in Equation~\ref{eq1} is rewritten explicitly as $\dot{\theta} = a(\theta) + b(\theta)u$ and $L_a, L_b$ are the total Lie derivatives with respect to $a, b$, the state matrix and the input matrix, respectively.
This brings us closed to a very important result, namely
\begin{align}
\dot{\theta}(t) = a(t, \theta) + b(t, \theta)u(t), \sigma = \sigma(t, \theta) &\\
\sigma^{(l)}(t) = h(t, \theta) + g(t, \theta)u(t)
\label{eq13}
\end{align}
where $h(t, \theta) = \sigma^{(l)}|_{u=0}$ and $g(t, \theta) = \frac{\partial}{\partial u}\sigma^{(l)} ~= 0$ supposing that for $K_m, K_M, C > 0$ 
\begin{align}
0 < K_m \leq \frac{\partial}{\partial u}\sigma^{(l)} \leq K_M &\\
\| \sigma^{(l)} |_{u = 0} \| \leq C
\label{eq14}
\end{align}
is always true (at least) locally. The control law $u$ would satisfy $u = \Psi(\sigma, \dot{\sigma}, \ddot{\sigma}, ..., \sigma^{(l)})$ hence for $\sigma = 0$ then $\Psi(0, 0, 0, ..., 0) = -\frac{h(t, \theta)}{g(t, \theta)}$. In other words we can write $\sigma^{(l)} \in [-C, C] + [K_m, K_M]u$ as a convex, semi-continuous function. This ensures that the closed-loop system trajectories in the plane $(\dot{\sigma}, \sigma)$ are confined between the trajectories of 
\begin{align}
\ddot{\sigma} = \pm C + K_m\Psi(\dot{\sigma}, \sigma) &\\
\ddot{\sigma} = \pm C + K_M\Psi(\dot{\sigma}, \sigma)
\label{eq15}
\end{align}
which provides the (homogeneous) control law $u$ as 
\begin{align}
u = -\alpha sat(\frac{\dot{\sigma}}{\dot{\Phi}} - \beta\frac{\sigma\sqrt{\sigma}}{\Phi^2})
\label{eq16}
\end{align}
with $\alpha K_m - C > \frac{\beta^2}{2}$. This control law has two components: namely a piece-wise discontinuous component $\beta\frac{\sigma\sqrt{\sigma}}{\Phi^2}$ and an energy dissipation component $\frac{\dot{\sigma}}{\dot{\Phi}}$ which determine the finite convergence time to the sliding manifold. This combination allows for asymptotic convergence under "unmatched" perturbations and uncertainty, as analytically proven in the work of \cite{kochetkov2022new}. Moreover, this allows the closed loop system to handle unstructured uncertainty (i.e. frequency domain disturbances typical for oscillator-based dynamics) beyond the typical robust formulation using $H\infty$ weighting, as described in \cite{owen2020unstructured}. Supported by the analysis of \cite{jayasuriya2020frequency}, the control law in Equation~\ref{eq16} can handle parametric uncertainty at low frequency and unstructured uncertainty at high frequency.

Finally, a further distinguishing property of a sliding mode is that it can become insensitive to changes in the dynamic properties of the controlled system under certain situations (i.e. traffic disruptions). It is critical to note that, unlike continuous systems with non-measurable disturbances where the invariance constraints need the employment of indefinitely high gains, the same effect is achieved in discontinuous systems by utilizing finite control actions.

\subsubsection{Control Synthesis}

\begin{figure}[!h]
\centering
\includegraphics [scale=0.32] {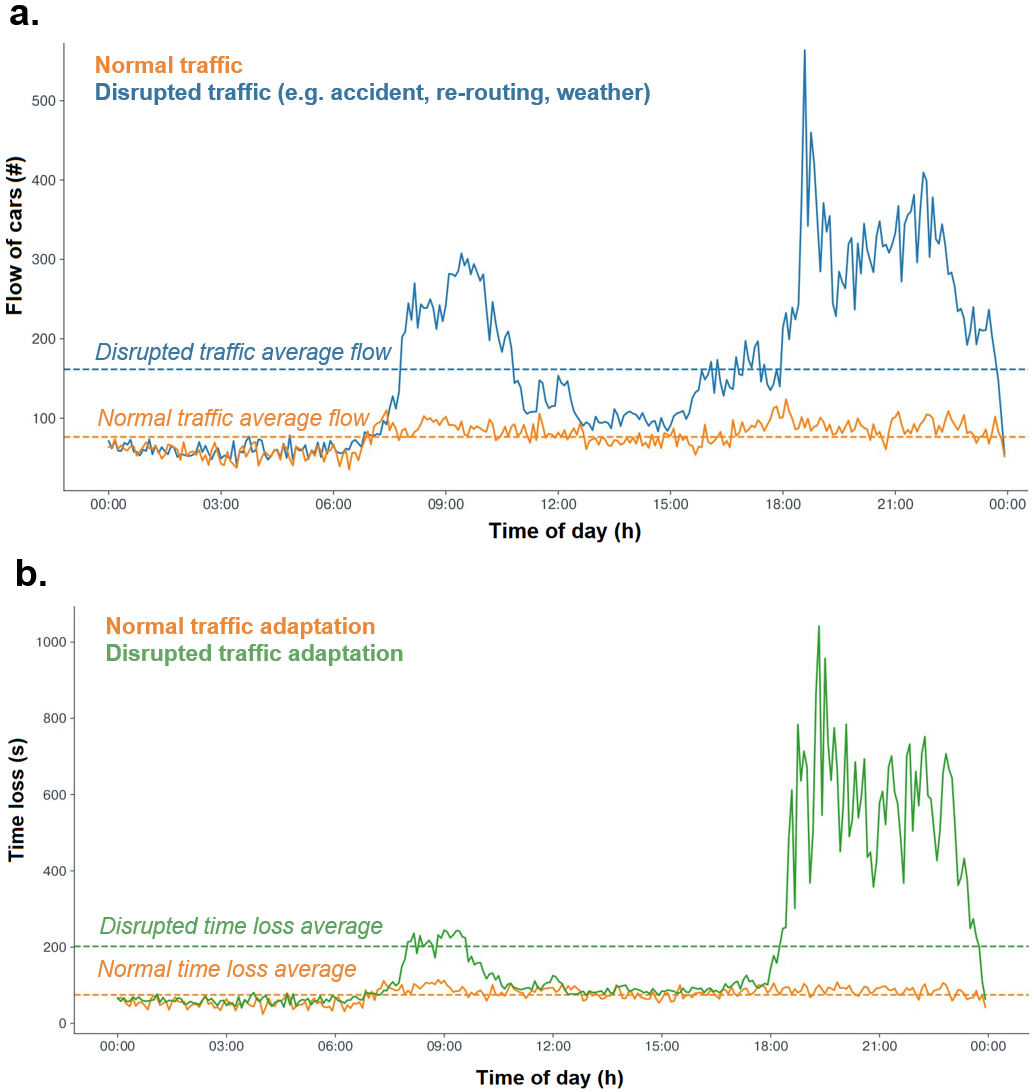}
\caption{Traffic dynamics in a daily urban scenario controlled by the model in Equation~\ref{oscillator_eq}. a) The traffic flow profile over a day in a Chinese town (8 intersections) when considering the network of oscillators model. b) Time loss during normal traffic over a day in a Chinese town (8 intersections) when considering the network of oscillators model. Changes in scale that propagate in time when disruptions are present.} 
\label{fig8}
\end{figure}

The network dynamics of the antifragile control (Equation~\ref{oscillator_eq}), is judiciously parametrized to cope with the normal daily traffic profile. This is visible in Figure~\ref{fig8} where the model is able to keep the lost time through a single cross to an acceptable value, around $70 s$ (see Figure~\ref{fig8} b). In the case of traffic disruptions (e.g. accident, sport events, or adverse weather conditions), the system cannot capture the fast changing dynamics (i.e. steep derivatives) of the traffic flow (see Figure~\ref{fig8} a) and, hence, performs poorly, for instance in preserving an acceptable time loss (i.e. difference in the duration of a trip in the traffic free vs.~heavy traffic) over rush-hour (see Figure~\ref{fig8}b around 18:00). 
The example in Figure~\ref{fig8} illustrates a limitation of such dynamic networked models, namely handling uncertainty. Be it structured uncertainty (e.g. sub-optimal choice of the internal oscillator frequency $\omega$ or a sudden time varying topological coupling $A_{ij}$ through trajectory re-routing) or unstructured uncertainty (e.g. un-modelled dynamics through the single use of $\dot{\theta}(t)$ and neglecting rate of change given by the Laplace operator $\ddot{\theta}(t)$), the system in Equation~\ref{oscillator_eq} is unable to converge to a satisfactory solution given input $k$ and coupling constraints.

To address this challenge, we extend Equation~\ref{oscillator_eq} with an antifragile control law. We choose to systematically maintain stability of the oscillatory dynamics by using a control approach which ensures consistent performance in the face of uncertainties. We use the result we obtained in Equation~\ref{eq16} and particularize for our the main state change in Equation~\ref{oscillator_eq}.

Given that the control law $u_i$ is given by Equation~\ref{eq16} and by rewriting the system as $\dot{\theta_i} = a_i(\theta_i) + b_i(\theta_i)u_i$, we then have
\begin{align}
u_i = -\alpha sat(\frac{\dot{\sigma}}{\dot{\Phi}} - \beta\frac{\sigma\sqrt{\sigma}}{\Phi^2}) &\\
a_i = \omega_i(t) + k_i(t)\sum_{j=1}^N{A_{ij}sin(\theta_j(t) - \theta_i(t))} + F_i sin(\theta^*(t) - \theta_i(t))
\label{eq17}
\end{align}
with the control gain $b_i$ given by
\begin{equation}
\begin{aligned}
&b_i(t) = \delta_{1}\int_{0}^{t} \hat{s_i}(\tau)d\tau \\
&\frac{d\hat{s_i}(t)}{dt} = \delta_{2}(\sum_{i,j}(\hat{s_j}(t)-\hat{s_i}(t)) + s_i(t)) \\
&\frac{ds_i(t)}{dt} = \delta_{3}\sum_{j}(s_j(t)-\frac{d\hat{s_i}(t)}{dt}) - \sign(\hat{s_i}(t))\frac{d^2\theta_i(t)}{dt^2} \\
&0<\delta_{1}<\delta_{2}<\delta_{3}<1
\end{aligned}
\label{eq18}
\end{equation}
where:\\
$s_i(t)$ - the surplus energy of traffic light $i$ oscillator \\
$\hat{s_i}(t)$ - the estimated surplus energy of traffic light $i$ oscillator \\
This choice, at its core, captures and controls the impact of higher-order motion (i.e. second derivative) through a high-frequency switching of the control law towards synchronization. Such a discontinuous antifragile control "drives", through a regularizing control law term $u_i(t)$, the coupled dynamics of the oscillators towards a desired dynamics (i.e. sliding surface $\sigma$).

\begin{figure}[!h]
\centering
\includegraphics [scale=0.32] {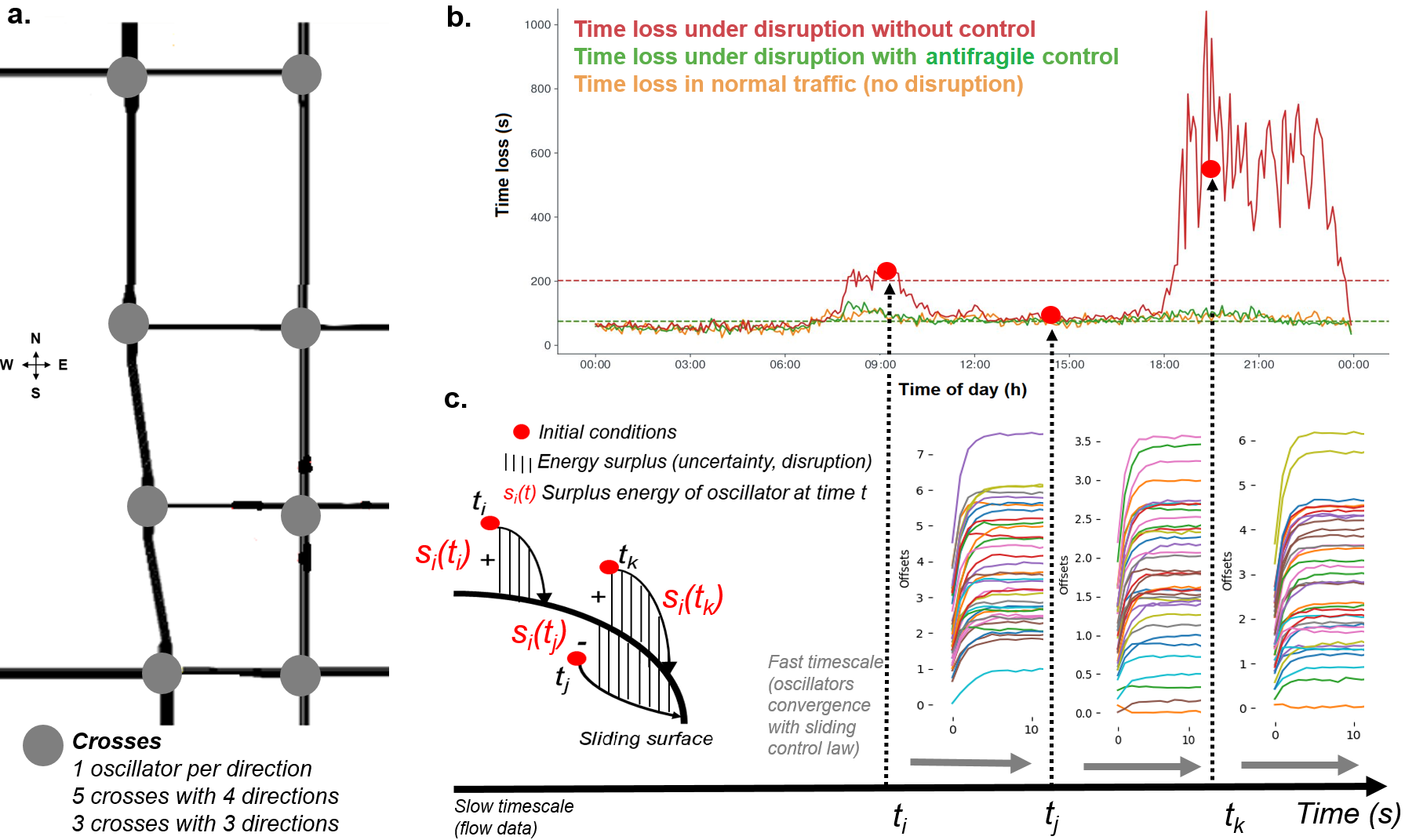}
\caption{Antifragile Control for oscillator-based road traffic model. In each cross of the considered network, an oscillator (given by Equation~\ref{oscillator_eq}) controls the green time for each direction (i.e. N, W, S, E). A full day of traffic is used to evaluate the performance w/wo disruptions and w/wo control. Each sample of measured traffic data is fed to the model which converges (on a fast timescale) to an estimate of green time given all spatial and temporal interactions with the other oscillators. a) Road network geometry and associated oscillators. b) Comparative time loss analysis of oscillator-based network model without disruptions, without control, and with antifragile control. c) Internal dynamics of the sliding modes and convergence dynamics of the oscillators (i.e. convergence). } 
\label{fig9}
\end{figure}

The goal of the regularizing sliding mode control law $u_i$ is to "push" the network of coupled oscillators, with a step size of $\delta$, towards a dynamics which accommodates the disruptions in the flow of cars $k$ (i.e. captured by $\ddot{\theta_i}(t)$). Intuitively, this assumes that the controller captures the second-order motion (i.e. $\ddot{\theta_i}(t)$) of the oscillator and compensates for it asymptotically until the surface $\sigma$ is reached. This assumes, in first instance, choosing an appropriate sliding surface $\sigma$ that minimizes the energy surplus $s_i(t)$ as illustrated in Figure\ref{fig8} c. Following Equation~\ref{eq18}, the regularizing control law $u_i(t)$ applied to oscillator $i$ is the area under the curve (i.e. the integral) of the estimated energy surplus, depicted in Figure~\ref{fig9} c. Interestingly, the (estimated) surplus energy, which keeps oscillator $i$ away from the desired antifragile dynamics $\hat{s_i}(t)$ depends on the local oscillators interaction $\sum_{i,j}(\hat{s_j}(t)-\hat{s_i}(t))$ and the actual surplus energy. The change in surplus energy is the actual dynamics of convergence to the sliding surface and is based on the cumulative impact of neighboring oscillators $\sum_{j}s_j(t)$ and the Laplacean of the green time $\ddot{\theta_i}(t)$ weighted by the direction of the convergence $\sign(\hat{s_i}(t))$.
The property of insensitivity of sliding surface in Equation~\ref{eq18} to the oscillatory dynamics \footnote{For a thorough analysis of sliding modes invariance see \cite{utkin2008sliding}.} is utilized to control the reaction of the network of coupled oscillators to uncertainty. We realized this practically by adding the regularizing term $u_i(t)$ in the local dynamics of each oscillator described by Equation~\ref{oscillator_eq}.
To get a better understanding of Equation~\ref{eq18}, we now exemplify, in Figure~\ref{fig9}, the impact the sliding mode controller has upon the dynamics of a road network when facing traffic disruptions from a real scenario (details about the data is provided in the Experiments and Results section). We consider a region composed of 8 crosses and $N=29$ oscillators as described in Figure~\ref{fig9} a.
In our case, the network of coupled oscillators is a system with discontinuous control (i.e. the control law $u_i(t)$ uses the sign of the energy surplus to drive the system towards the antifragile dynamics). Basically, as shown in Figure~\ref{fig9} b, c, given each sample of flow data $k_i(t_i ... t_j ... t_k)$ (from the road sensors) there is a fast convergence time-scale which allows the oscillators to reach steady state. This state is reached under sliding mode control by compensating for the disruptions in the traffic flow modelled by the second-order motion $\ddot{\theta_i}(t)$. The stationary state is subsequently probed for the actual phase duration, relative to each coupled oscillator by solving ${\arg \max} \{ \rho(t) > \tau \}$ where $\rho_{ij}(t) = cos(\theta_i(t) - \theta_j(t))$ and $0 < \tau < 1$. Due to the fast changes occurring during disruptions (see Figure~\ref{fig9} b - rush hour around 6 PM), in the slow time-scale of traffic flow (i.e. sensory data), the network of coupled oscillators benefits from the sliding mode control law to compensate for the abrupt changes and to reach consensus, as shown in Figure~\ref{fig9} c - right panel. This consensus state describes the point when the system dynamics reached the sliding surface, in other words when the magnitude of the surplus energy decayed at a finite rate over the finite time interval (i.e. fast timescale in Figure~\ref{fig9} c - left panel).
The regularization approach we propose has a simple physical interpretation. Uncertainty in the system behavior in the face of disruptions appears because the motion equations of the dynamics in Equation~\ref{oscillator_eq} are an ideal system model. Non-ideal factors such as un-modelled dynamics and sub-optimal parameter selection are neglected in the ideal model. But, incorporating them into the system model eliminates ambiguity in the system behavior which "slides" to an antifragile dynamics.

\section{Experiments and Results}

The experiments and evaluation use the SUMMER-MUSTARD (Summer season Multi-cross Urban Signalized Traffic Aggregated Region Dataset) real-world dataset, which contains 59 days of real urban road traffic data from 8 crosses in a city in China. The road network layout underlying is depicted in Figure~\ref{fig7} a. In order to perform experiments and evaluate the system, we used the real-world traffic flows in the Simulator for Urban Mobility (SUMO)\cite{SUMO2018}. This realistic vehicular simulator generates routes, vehicles, and traffic light signals that reproduce the real car flows in the dataset. 

In our experiments, we comparatively evaluated the adaptive behavior of the antifragile control and relevant state-of-the-art approaches against the static traffic planning (i.e. police parametrized phases) used as baseline. We performed an extensive battery of experiments starting from the real-world traffic flows recorded over the 8 crosses in the SUMMER-MUSTARD dataset. In order to evaluate the adaptation and antifragility capabilities, we systematically introduced progressive magnitude disruptions over the initial 59 days of traffic flow data. Disruptions, such as accidents and adverse weather determine a decrease in the velocity which might create jams (see Figure~\ref{fig2}). Additionally, special activities such as sport events or beginning/end of holidays increase the flow magnitude. Such degenerated traffic conditions might happen due to non-recurrent events such as accidents, adverse weather or special events, such as football matches. Using the real-world flow and SUMO, we reproduce the traffic flow behavior when disruption occurs starting from normal traffic flow data by reflecting the disruption effect on vehicles speed and/or network capacity and demand. We sweep the disruption magnitude from normal traffic up to 5 levels of disruption reflected over all the 8 crosses over the entire day.

The evaluated control systems are the following:
\begin{itemize}
\item BASELINE - An optimized static traffic planning that uses pre-stored timing plans computed offline using historic data in the real-world. 
\item OPTIMAL - An optimal control method based on MILP phase plan optimization implementation inspired from \cite{ouyang2020large}.
\item ROBUST - A basic implementation of a robust control based on a network of Kuramoto oscillators (\cite{strogatz2000kuramoto}) for each direction in the road network cross inspired by \cite{agrawal1998non} and \cite{owen2020unstructured}.
\item ANTIFRAGILE - The antifragile control with the regularizing sliding mode control law $u$.
\end{itemize}

\subsection{Evaluation of the phase calculation accuracy}

For the evaluation of the different approaches for phase duration computation (i.e. BASELINE, OPTIMAL, ROBUST, and the ANTIFRAGILE control), we followed the next procedure:
\begin{itemize}
\item Read relevant data from simulation experiment (without disruptions and with 5 levels of progressive disruptions) for each of the systems.
\item Compute relevant traffic aggregation metrics (i.e. average time loss, average speed, and average waiting time).
\item Rank experiments depending on performance.
\item Perform statistical tests (i.e. a combination of omnibus ANOVA and posthoc pairwise T-test with a significance $p = 0.05$) and adjust ranking depending on significance.
\item Evaluate best algorithms depending on ranking for subsets of relevant metrics (i.e. the metrics with significant difference).
\end{itemize}
Our evaluation results are given in Table~\ref{perf-table} where each of the approaches is ranked across the disruption magnitude scale (no disruption to maximum disruption) over the specific metrics (i.e. average time loss, and average speed, and waiting time, respectively). For flow magnitude disruptions, the level of disruption (i.e. 1.1 ... 1.5) is a factor used to adjust the number of vehicles or the speed of vehicles (i.e. for adverse weather) during the disruption. The evaluation was performed on the entire dataset containing recorded traffic flows over 59 days from 8 crosses.
\begin{table*}\centering
\ra{1.0}
\begin{tabular}{@{}rrrrrrrrrrrrcrrrrrrrrrcrrrrrrrr@{}} & \phantom{rrrrrrrrr}& \\
\midrule 
Control System/\\Disruption level & normal flow & 1.1 & 1.2 & 1.3 & 1.4 & 1.5 \\ \midrule
\textit{Average time loss(s)}\\
BASELINE & 102.535 & 114.600 & 136.229 & 241.383 & 197.399 & 202.113\\
OPTIMAL & 151.281 & 153.781 & 203.301 & 309.671 & 223.017 & 257.464\\
ROBUST & 131.468 & 161.871 & 203.301 & 309.671 & 199.797 & 497.124 \\
ANTIFRAGILE  & 85.726 & 88.326 & 89.726 & 84.165 & 89.889 & 84.291\\
\midrule
\textit{Average speed*}\\
BASELINE & 5.81 & 5.67 & 5.46 & 5.02 & 4.94 & 4.75 \\
OPTIMAL & 5.97 & 5.87 & 5.46 & 5.03 & 4.92 & 4.61\\
ROBUST & 5.94 & 5.81 & 5.46 & 5.02 & 5.29 & 4.54\\
ANTIFRAGILE   & 5.98 & 5.97 & 5.94 & 5.18 & 5.07 & 5.15 \\
\midrule
\textit{Waiting time(s)}\\
BASELINE & 164.5 & 185.3 & 222.8 & 294.5 & 325.9 & 351.3 \\
OPTIMAL & 148.7 & 148.7 & 212.8 & 234.5 & 293.2 & 372.9\\
ROBUST & 115.7 & 142.2 & 215.8 & 286.5 & 208.5 & 418.3\\
ANTIFRAGILE  & 128.7 & 145.7 & 148.8 & 159.2 & 162.4 & 158.7\\
\bottomrule
\end{tabular}
\caption{Performance evaluation for the different phase duration calculation methods. *Average speed calculated as the ratio between distance traveled and time of travel.}
\label{perf-table}
\end{table*} 
The exhaustive experiments and evaluation in Table~\ref{perf-table} demonstrate where our approach excels and where it fails to provide the best phase duration calculation. The chosen evaluation metrics reflect the overall performance (i.e. over multiple days) with respect to the most significant traffic metrics given the phase duration value computed by each of the systems.

The previous analysis is supported by the normalized ranking over the entire SUMMER-MUSTARD dataset in Figure~\ref{fig6}, where we provide a condensed visual representation of each system's performance.
Here, we can see that if we consider the average time loss, the baseline performs worst due to its pre-defined timings and inability to adapt to unexpected disruptions during the daily traffic profile. 
At the other end of the ranking, both implementations of ROBUST the ANTIFRAGILE control provide minimal waiting time capturing fast and steep changes in the disrupted flows. 
Due to their similar core modelling and dynamics, the ANTIFRAGILE control systems and ROBUST tend to provide similar performance, with a relative improvement on the antifragile control side in terms of duration, waiting time, and speed metrics. This is due to its spatio-temporal extension beyond the basic oscillator model that can capture also the spatial contributions of adjacent flows beyond their temporal regularities when computing the phase duration. 
Finally, the ROBUST the ANTIFRAGILE control system provides overall superior performance through its discontinuous sliding mode control law that captures the deviation of the dynamics in the presence of disruptions and compensates robustly for their impact on the oscillator convergence (see Figure~\ref{fig7}).
\begin{figure}[!h]
\centering
\includegraphics [scale=0.5] {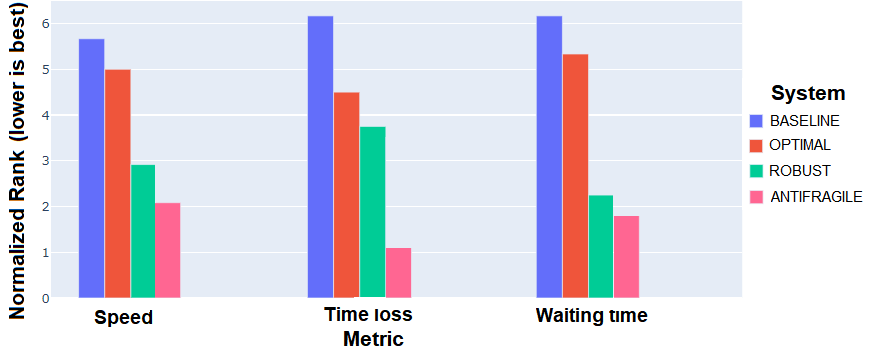}
\caption{Control Systems Ranking on All Metrics and Entire Dataset (8 crosses over 59 days).} 
\label{fig10}
\end{figure}

\subsection{Evaluation of the run-time}

In terms of run-time, the adaptive methods (i.e. OPTIMAL, ROBUST, and ANTIFRAGILE) provide different levels of performance, mainly due to the modelling and optimization types they use. The baseline is excluded as it is just the static optimized plans allocation for the real traffic setting in SUMMER-MUSTARD, basically, a simple value recall from a look-up-table. We measured the time needed by each of the evaluated adaptive systems to provide a phase duration estimate after a sensory sample (i.e. one sensory reading of traffic flow data). As mentioned, each system uses a different computational approach: the optimal uses a solver that implements an LP-based branch-and-bound algorithm, the ROBUST uses a Runge-Kutta 45 ODE solver, and the ANTIFRAGILE control is implemented using Runge-Kutta 45 ODE solver. The evaluation is given in Table 2 where the average value over the entire range of traffic conditions (normal and disruptions) is considered. The experimental setup for our experiments used 3 machines, each with 24 CPU cores and 132 GB RAM, and Apache Flink for stream processing and cluster management.
\begin{table}\centering
\begin{tabular}{c c c}
Model & Single cross & Region (8 crosses)  \\
\hline
OPTIMAL  & 0.0510 & 0.3930 \\ [1ex]
ROBUST & 0.0568 & 0.4544 \\ [1ex]
\textbf{ANTFRAGILE} & \textbf{0.0071} & \textbf{0.0426} \\
\hline
\label{runtime-table}
\end{tabular}
\caption{Adaptive phase duration calculation run-time evaluation.}
\end{table}
As expected, at the level of a single intersection optimization, OPTIMAL and ROBUST controls lie in the same range, providing a new phase duration value after 50 ms. At the region level, considering all 8 crosses, the run-time increases with an order of magnitude, with OPTIMAL overtaking the ROBUST due to OPTIMAL's constraint optimization efficiency at scale and the similar computations of ROBUST control. The fastest approach, both as single cross and regions level, is the ANTIFRAGILE control. With more than 80\% run-time improvement both at single cross-level and regional-level, the ANTIFRAGILE control excels due to its efficient and simple computations.

\section{Discussion}

Traditionally, phase duration optimization for coordinated traffic signals is based on average travel times between intersections and average traffic volumes at each intersection. We now present our main take-away from the study and the new instantiation of antifragile control.

\subsection{Modelling}
Our study introduces a novel instantiation of antifragile control for a novel model of road traffic phase duration optimization applicable to any road traffic layout, scale, and architecture (i.e. number of lanes per direction etc.). More precisely, using an oscillator-based model \cite{strogatz2000kuramoto} of the traffic flow dynamics in large signalized road networks, the proposed closed-loop control system exploits the periodic nature of the traffic signal circular phasing similar to \cite{akbas2005dynamic,fang2013matsuoka}. Our instantiation of the antifragile control goes beyond optimal and robust control approaches by considering a weighted external perturbation (e.g. cycle time reference weight $F$), flow modulation $k$, and a spatial topology weight $A$ which are regularized by a sliding mode control law $u$.  Such a modelling and control approach adapts to unpredictable disruptions in traffic flows (e.g. accidents, re-routing, adverse weather conditions) up to a certain extent, where the dynamics of the disruption doesn't perturb the self-organization of the coupled oscillators under the sliding dynamics of the closed-loop dynamics given the control law. 

\subsection{Antifragile control}
In reality, the "steep derivatives" of traffic flows do not allow the optimal and robust control to converge to the best phase duration value. In order to achieve high-performance (i.e. minimizing metrics such as time loss or maximizing average speed), we synthesized an antifragile controller for the network of oscillators model. Such a controller "pushes" the perturbed dynamics under disruptions towards a dynamics that "drive" the coupled oscillators network towards the optimal phase. This way, the ANTIFRAGILE control system is capable to solve local and global traffic dynamics by exploiting the coupling among different oscillators describing traffic periodicity under disruptions.
The proposed optimal control system in \cite{ouyang2020large} - termed OPTIMAL in our experiments - used exact mathematical programming techniques (i.e. mixed-integer linear programming) for optimizing the control of traffic signals and has shown only limited adaptation capabilities.
As our evaluation shows, OPTIMAL performs well, for instance in minimizing route length, but carries a large computational cost due to the optimization process that needs to iterate to convergence.
Finally, we can see that ANTIFRAGILE control excels in both Speed and Time Loss metrics, depicting the fact the control law choice based on the second-order effects of the signal re-computation can capture such dynamics of the closed-loop system. Additionally, the results clearly indicate the over-performance of ANTIFRAGILE over ROBUST and OPTIMAL in terms of transient parameters and steady state error. In our experiments we observed that OPTIMAL control tried to optimize the performance metrics over the given span of time but was very sensitive to the oscillator-based network model parameters, while ROBUST control tried to optimize the stability and quality of the response given mild levels of uncertainty (i.e. disruptions) in the oscillator-based model. The ANTIFRAGILE uses variable structure control to handle both structured and unstructured uncertainty in the model by synthesizing a second-order effects-aware control pushing the system to the desired antifragile region of the state space.

\section{Conclusion}

Traffic control is a multi-dimensional problem to be optimized under deep uncertainty. Modelling traffic dynamics is fundamental for traffic control. Aiming at capturing the periodic nature of traffic, we propose an instantiation of antifragile control for a model of urban road traffic using a network of oscillators capturing the spatial and temporal interactions among different crosses in a traffic network. In order to adaptively cope with unexpected traffic flow disruptions, the antifragile control uses a second-order sliding mode controller that strengthens its adaptation capabilities towards global consensus under high-magnitude disruptions and uncertainty. The system is lightweight and exploits the coupling interactions among different controlled oscillators. 
Our evaluation of the system on real-world data and against state-of-the-art methods demonstrates the advantages the antifragile control brings. From capturing the periodic dynamics of traffic phasing, to embedding the spatial correlation among traffic flow along its temporal dimensions, and up to robustly adapting to unexpected traffic disruptions, the antifragile control stands out as a flexible solution for green time phase duration calculation. We believe that antifragile control is a strong candidate for actual real-world deployment. 

% \bibliography{bibliography}

\end{document}